\documentclass[mathleft,fleqn,%
]{an} 
\usepackage{longtable}
\usepackage{tablefootnote}
\usepackage{graphicx}
\usepackage[varg]{txfonts}
\usepackage{apacite}
\usepackage[authoryear]{natbib}
\begin{document}

\Pagespan{1}{}
\Yearpublication{2018}%
\Yearsubmission{2018}%
\Month{0}%
\Volume{999}%
\Issue{0}%
\DOI{asna.201400000}%

\title{Chemical composition of post-AGB star candidates}

\author{R. E. Molina \inst{1}\fnmsep\thanks{Corresponding author:
        {rmolina@unet.edu.ve}}
C. B. Pereira\inst{2} \and A. Arellano Ferro\inst{3}
}
\titlerunning{PAGB atmospheric abundances}
\authorrunning{R.E. Molina et al.}
\institute{
Laboratorio de Investigaci\'on en F\'isica Aplicada y
Computacional, Universidad Nacional Experimental del T\'achira,  Venezuela.
\and 
Observat\'orio Nacional/MCTI, Rua Gen. Jos\'e Cristino 77, Sao Cristovao, Rio
de Janeiro, CEP 20921-400, Brazil.
\and
Instituto de
Astronom\'ia, Universidad Nacional Aut\'onoma de M\'exico, Ciudad
de M\'exico, CP 04510, M\'exico.}

\received{XXXX}
\accepted{Jan 16, 2019}
\publonline{XXXX}


\abstract{We present a high resolution detailed abundance
analysis for a sample of six post-AGB candidate stars, five of
them had not been studied spectroscopically in the optical
region. All the analyzed objects are IRAS sources identified as
possible post-AGB on the two-colours IRAS diagram. We find three
objects with clear signs of evolved stars; 
IRAS 05338-3051 shows abundances similar to the RV Tauri
V453 Oph; the lower-luminosity stars IRAS\,18025\,-\,3906 is
O-rich without s-process enrichment and IRAS\,18386\,-\,1253
shows a moderate selective depletion of refractory elements
generally seen in post-AGB stars, which show mid-IR excess; they
may be evolved post-RGB objects, in which case
these would be the first Galactic counterparts of post-RGB
objects observed in the Large and Small Magellanic Clouds (Kamath
et al. 2014, 2015). On the other hand, IRAS\,02528\,+\,4350
seems to be a moderately metal-poor young object and
IRAS\,20259\,+\,4206 also seems to be a young object showing
carbon deficiency; however, an analysis with better spectra
might be in order to clarify its evolutionary state. Finally,
our abundances calculations for the binary post-AGB star
IRAS\,17279\,-\,1119 are found in good agreement with those of
De Smedt et al. (2016).} 
  
 \maketitle

\section{Introduction}
\label{sec:introd}

Post-AGB stars (PAGBs) are luminous objects of low-to intermediate-mass (0.8--8.0 M$_{\odot}$) in the late 
stages of their evolution. These objects are located in the intermediate evolutionary stages between the end
of the AGB phase and the beginning of the planetary nebulae (PN). 
They display diverse chemical atmospheric
compositions as they 
are determined by nucleosynthesis and mixing processes and by extensive mass loss
occurring 
during the AGB phase of evolution \citep []{IbenRenzini1983, MarigoEtAl2013, KarakasEtAl2014}. The structure and evolution of low and intermediate mass stars
during 
the AGB and the PAGB phases has been described in detail in numerous papers 
\citep[e.g.][]{VassiliadisWood1993, Bloecker1995, BussoEtAl1999, 
Herwig2005}. It is clear now that severe mass loss at the end of AGB leads to
enrichment of the interstellar medium in carbon, nitrogen and s-process elements
\citep[e.g.] [] 
{IbenTruran1978, DrayEtAl2003, Karakas2010}. Hence, AGB stars play a very
important 
role in the chemical evolution of the galactic interstellar medium.

In massive AGB stars, heavy mass loss produces a dense circumstellar shell that obscures
the 
central regions whereas the central star may remain at least partially visible. The
dust in the circumstellar shells is heated by stellar radiation, and causes an
infrared (IR) excess. 
Hence, the optically visible galactic PAGBs generally exhibit two peaks in their spectral 
energy distribution (SED) and could be sepparated accordingly into two groups
of SEDs: disk-souces \cite [] {DeRuyterEtAl2006, GielenEtAl2009, VanWinckelEtAl2009} and 
shell-sources \cite [] {VanWinckel2003}. The first group
displays SEDs with a broad IR excess starting in the near-IR region, which point to the presence of hot dust near
the central star (these objects are associated with binary stars). The second group contains PAGBs with a 
double-peaked SED where the first peak corresponds to the photospheric flux and the second peak represents the IR
emission of an expanding dusty envelope (these objects are single stars).

Some subgroups of PAGBs with IR-excess may be identified according to their C/O ratio being either greater 
or smaller than one. A good account of the progress in our understanding 
of PAGBs can be found in the works of \cite{VanWinckel2003}, \cite{GarciaLario2006} 
and \cite{Giridhar2011}. PAGBs in LMC and SMC have been studied and summarized by \cite{VanAarleEtAl2011}, 
\cite{VanAarleEtAl2013} and \cite{KamathEtAl2014}.
Their known distances allow good estimates of their luminosity and mass. From their
chemical 
composition PAGBs can also be distinguished for example by the s-process
enhancement \citep[e.g.] [] {ReyniersEtAl2007} and by the depletion of condensable elements
\citep[e.g.] [] {GielenEtAl2009}. 
For a more recent summary of galactic and extra-galactic PAGBs see \cite {DeSmedtEtAl2016} and
the 
references therein. In these studies some inadequacies of theoretical AGB models in
explaining the 
observed low C/O ratios in s-process enriched stars \citep [] {DeSmedtEtAl2012, DeSmedtEtAl2014, VanAarleEtAl2013} 
and the low upper limit of Pb are remarked \citep [] {DeSmedtEtAl2014}. 

PAGBs with C/O $>$ 1,  e.g. HD\,56126, HD\,187785 and IRAS\,06530\,-\,0213 display a
considerable enhancement of C and s-process elements \citep[e.g.] [] {VanWinckelEtAl1997, ReddyEtAl2002, 
ReyniersEtAl2002, ReyniersEtAl2004}. PAGBs exhibiting s-process
elements
caused by the third dredge-up (TDUP) make only a
small fraction of the known PAGBs and they are generally single stars. 

PAGBs with C/O $<$ 1, e.g. 89 Her, HD\,161796, HD\,133656, SAO\,239853, etc, are
O-rich and do not display s-process enhancement, whilst many PAGBs with C/O $\sim$ 1
e.g.
BD\,+\,39\,4926, HR\,4049, HD\,44179, HD\,46703, HD\,52961 \citep []{VanWinckelEtAl1995, VanWinckel2003}
show abundance peculiarities caused by selective removal of condensable elements. This
selective
removal of refractory 
elements has been attributed to the presence of a stellar companion and the
circumstellar envelope 
that provides the site for dust-gas separation to take place. Re-accretion of clean
gas from the 
circumbinary shell to the photosphere, while the dust remains stable in the
circumstellar shell 
(due to large radiation pressure on them), results in anomalous chemical composition
exhibited by 
these objects. This scenario, proposed by \cite {VennLambert1990}, \cite {Bond1991} and
elaborated by \cite {WatersEtAl1992}, has been gaining support through the observed depletions of
condensable elements in a large number of PAGBs, including RV Tauri stars \citep{ GiridharEtAl1994,  
VanWinckelEtAl1995, VanWinckelEtAl1997,  GonzalezEtAl1997,  GiridharEtAl2000,  GiridharEtAl2005}.
Studies of extended samples of PAGBs, including RV Tauri stars with dusty disk, have
been carried out by \cite {MaasEtAl2002} and  \cite {DeRuyterEtAl2006} and the
chemical anomalies
arising due to depletion of condensable elements is observed at various degrees, which
further supports
the above mentioned hypothesis. Binary companions are detected for a large number of
PAGBs showing 
depletions and a sizable fraction of them being RV Tauri stars. The presence of a dusty disk,
generally inferred from the shape of their SED, is also supported by mid-IR
interferometry \citep [] {DerooEtAl2006}, which points towards the existence of compact
disks. Through time series radial velocity monitoring, 
\cite {VanWinckelEtAl2009}
computed the orbital
elements of many PAGB binaries showing removal refractory elements depletions. 

A detailed study of a set of RV Tauri and PAGBs based on the analysis of their SED
(shell and disk 
sources) carried out by \cite {GezerEtAl2015}, has shown that there is a
correspondence between the 
properties of the SED and binarity. These authors noted a complex relationship between
the IR properties 
and chemical anomaly and they reported that all confirmed binaries are disk sources
but not all of
them show depletion patterns.

However, among depleted sources there are some objects with apparently no IR excess.
An example is the heavily depleted object BD\,+\,39\,4926. The depletion effect could be explained by its binary nature in spite of a lack of detected IR 
excess. According to \cite {GezerEtAl2015} the depleted photosphere remains chemically peculiar 
for a longer time than the IR excess remains detectable. 

However, \cite {VennEtAl2014}, using new photometry 
data taken from WISE database, have demonstrated that BD\,+\,39\,4926 has a mild-IR
excess beyond 10 microns. 
The radial velocity curve of BD\,+\,39\,4926 together with orbital parameters was
obtained by \cite {GezerEtAl2015}.

The IRAS two colour diagram and the $J-K$ versus $H-K$ diagram have been very useful
tools in identifying new source candidates to PAGBs and PN
\citep [] {PreiteMartinez1988, VanderVeenHabing1988, GarciaLarioEtAl1997}. However, the confirmation 
of these sources as PAGBs, requires detailed chemical composition 
analysis from high-resolution spectroscopy. The extensive IRAS source list studied by
\cite {GarciaLarioEtAl1997} and its refinement in the optical from low-resolution spectroscopy 
by \cite {SuarezEtAl2006} and \cite {PereiraMiranda2007}, have contributed to the
identification of new candidate objects to PAGBs. More recently, \cite {SzczerbaEtAl2017} show 
the results of the search of new AGB (2,510) and PAGB (24,821) star candidates using 
the [3.4]-[12] vs [4.6]-[22] two colour diagram from ALLWISE photometry.

Although the number of known PAGBs has grown considerably, their chemical diversity,
particularly among 
objects of very similar atmospheric parameters, highlights the complexities involved
and our limited understanding 
of the processes operating during the relatively fast AGB and PAGB evolution. Hence,
identification, detailed
atmospheric analysis and SED properties of new PAGB stars is important in providing
observational constrains
to refine the models of PAGB evolution, including the influence of binary companions.

In continuation of our search and analysis of new candidates to PAGBs \citep [] {GiridharEtAl2010, MolinaEtAl2014}, 
in the present paper we carry out a high resolution detailed abundance analysis for a
sample of five candidate objects; IRAS\,02528\,+\,4350,  
IRAS\,05338\,-\,3051,  IRAS\,18025\,-\,3906, IRAS\,18386\,-\,1253
and IRAS\,20259\,+\,4206, selected via the IRAS two colour diagram, and of the binary post-AGB star IRAS 17279-1119.

This work is structured as follows; in Section 2  we describe the selection process of
our sample. In Section 3 we give 
a description of the observations. In Section 4 the initial atmospheric parameters
calculation is described.
In Section 5, the methodology towards the abundances calculation and their
uncertainties is presented. In Section 6, we report the 
individual abundances for each star in our sample. Section 7 is dedicated to
the discussion of the implications of the chemical 
results on the status of each objects, and finally we summarize our conclusions in
Section 8.

\section {THE SAMPLE}
\label{sec:sample}

\begin{figure}
\begin{center}
  \includegraphics[width=\columnwidth]{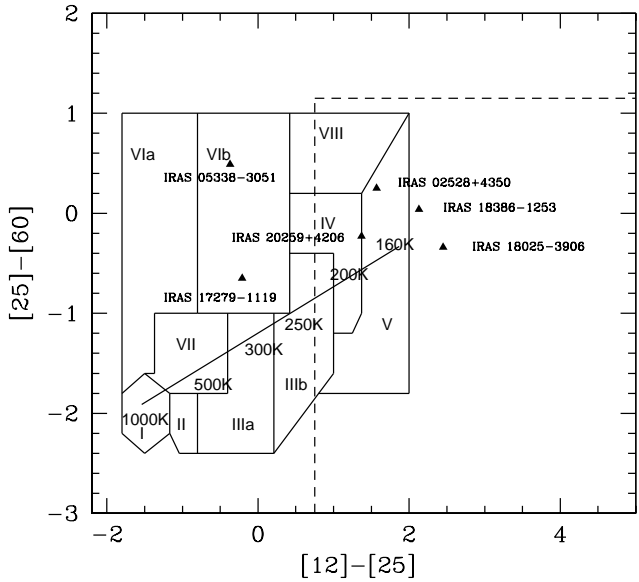}
  \caption{Location of the sample stars in the IRAS two colour diagram. The
dotted-line represents the
PNs and PAGBs region limited by \cite {GarciaLarioEtAl1997}. The different regions
represented
with roman numerals have been defined by \cite {VanderVeenHabing1988} and separate stars by 
the dust/gas ratio in the 
envelope. The black body curve is represented by the straight line.}
\label{fig:figure1}
\end{center}
\end{figure}

The IRAS two colour diagram \citep {VanderVeenHabing1988, GarciaLarioEtAl1997}
has been very useful in separating 
and identifying different types of luminous evolved stars, in particular those in the
post-AGB evolutionary phases 
(see Fig.~\ref{fig:figure1}). The present sample contains the stars
IRAS\,02528\,+\,4350, IRAS\,05338\,-\,3051,
IRAS\,17279\,-\,1119, IRAS\,18025\,-\,3906, IRAS\,18386\,-\,1253 and
IRAS\,20259\,+\,4206. Four of them IRAS\,02528\,+\,4350, 
IRAS\,18025\,-\,3906, IRAS\,18386\,-\,1253 and IRAS\,20259\,+\,4206 fall within the
region for PNs and PAGBs defined
by \cite {GarciaLarioEtAl1997}, 
and the other two IRAS\,05338\,-\,3051 and
IRAS\,17279\,-\,1119, fall in the VIb region
which, according to \cite {VanderVeenHabing1988} contains variable stars with
relatively hot dust closer to the star
and colder dust at larger distances. A brief review of the six objects is given below. 

IRAS\,02528\,+\,4350 (GLMP 34) is a high-latitude galactic object and was included in the sample of \cite {PreiteMartinez1988} as a PN candidate.  Preite-Mart\'inez 
derived a dust temperature of 132K and estimated a distance of 3.8 kpc. This object was included as an IRAS galaxy due to 
its IR-excess at 60$\mu$m band 
\citep [] {Cohen1992, NakanishiEtAl1997}. With
[12]\,-\,[25]$=+$1.57 and 
[25]\,-\,[60]$=+$0.25, it belongs to the zone of PAGBs defined by \cite {GarciaLarioEtAl1997}. \cite {SuarezEtAl2006}, 
using low-resolution spectra, classified it as a young object of the A0e spectral
type. \cite {SzczerbaEtAl2007} included it
in Toru\'n catalogue as possible a  PAGB object. \cite {VickersEtAl2015} proposed that it belongs to 
population of the thin disk, and adopting a core mass of 0.53$\pm$0.02 M$_{\odot}$, 
through the core
mass--luminosity relation of \cite {VassiliadisWood1993}, found a luminosity of
L$_{*}$\,$\approx$\,1700$\pm$750 L$_{\odot}$ 
and a distance of 4.34$\pm$0.98 kpc. 

IRAS\,05338\,-\,3051 (RV Col) is of high galactic latitude ($b$\,=\,$+$28.8), with spectral type G5 I \citep{Straizys1982}
and without an IR-excess observed.
It is classified as a semi-regular variable in 
the General Catalog of Variable Stars (GCVS). \cite {Hoffmeister1943} reports a period of
105.7 days. Light curves for different 
photometric bands indicate that the period is relatively stable with an amplitude of
$\sim$0.8 mag in the $V$-band, but according to \cite {Eggen1986} it displays
a "malformation" near 25 days after maximum light. 

The star has been monitored for about 9 years by the All-Sky Automated Survey (ASAS-3)
group. The light curve includes 600 observations
which exhibit a stable sinusoidal light variation with a variable amplitude of 0.5 to
1.0 mag. \cite {ArkhipovaEtAl2011} 
found two active periods in the light variations; 105.8$\pm$1.0 and 108.6$\pm$1.5
days with a maximum amplitude of 0.6 mag.
Although the amplitude varies, the period has remained
almost constant for decades \citep{ArkhipovaEtAl2011}.

\cite {TramsEtAl1991} pointed out that this object does not show infrared excesses and
suggested it to be a red giant of spectral 
type G. \cite {VanderVeenEtAl1993} studied the mass-loss rate and the expansion
velocity for a set of evolved stars, including 
IRAS\,05338\,-\,3051, however, the rotational molecular lines of $^{12}$CO J\,=\,1-0
and J\,=\,2-1 were not
sufficiently intense for a detailed analysis of this object.

IRAS\,05338\,-\,3051 is listed as a possible PAGBs into the Toru\'n catalogue of
post-AGB stars \citep [] {SzczerbaEtAl2007}. 
\cite {VickersEtAl2015} estimated a distance of 2.42$\pm$0.52 kpc from an adopted luminosity of
3500$\pm$1500 L$_{\odot}$. According to 
Vickers et al.\@ this object probably belongs to the old thin disk population. The
IRAS colours [12]\,-\,[25]\,=\, $-0.37$ and
[25]\,-\,[60]\,=\, $+0.51$ show that this object falls in the VIb region of
the colour-colour diagram of \cite {VanderVeenHabing1988}.

IRAS\,17279\,-\,1119 (HD\,158616) has been the subject of several abundance analysis
e.g. \cite{VanWinckel1997}, \cite{ArellanoEtAl2001},  
\cite{RaoEtAl2012}, \cite{DeSmedtEtAl2016}.  \cite{Bogaert1994} identified two possible 
periods between 61 and 93 days for this star. \cite{VanWinckel1997} points out that the
optical photometry shows similarity to RV 
Tauri stars. With [12]\,-\,[25]$= -0.21$ and [25]\,-\,[60]=$-0.65$, it
belongs to the zone VIb. This object has also been included in the Toru\'n Catalogue
of Galactic PAGB \citep []{SzczerbaEtAl2007} as a likely PAGBs.

An extensive monitoring of the radial velocity for about $\sim$6 yr \citep []{VanWinckelEtAl2010, GorlovaEtAl2015} led to determination of the radial velocity orbit. \cite{DeSmedtEtAl2016} also used the ASAS-3 photometry and found periods between 83 and 90 days. The SED constructed from different photometric sources by \cite{DeSmedtEtAl2016} (their figure 16) shows an infrared excess starting near 2 microns, which is indicative of the presence of warm dust at sublimation temperature near the central star.  It has been suggested by several workers \citep []{DeRuyterEtAl2006, DerooEtAl2007, GielenEtAl2011, HillenEtAl2013, HillenEtAl2014, BujarrabalEtAl2015, GezerEtAl2015} that this is caused by a stable compact circumbinary disk.

IRAS\,18025\,-\,3906 (GLPM 713) was included in the sample of \cite{PreiteMartinez1988} as
a PN candidate. However, from their analysis they concluded that
IRAS\,18025\,-\,3906 is not a PN but that the object appears to be a non-variable
OH/IR star and a proto-planetary nebulae (PPN) candidate. The author derived a dust temperature of 172\,K for
this object. \cite{HuEtAl1993} relate IRAS\,18025\,-\,3906 to the optical counterpart object with a G2\,I spectral type. In addition these authors detected OH-masers at 1612 and 1667 MHz and its SED shows a double-peak structure (see their Fig. 3, item \# 40).

\cite{SilvaEtAl1993} detected a double peaked OH-maser
emission at 1612 MHz with velocities of $-$103 and $-$130 km s$^{-1}$. They estimated a velocity (LSR) of $-$116 km s$^{-1}$,
an envelope expansion velocity of 13 km s$^{-1}$, a mass loss rate of M = 5.0 x
10$^{-6}$ M$_{\odot}$ yr$^{-1}$ and a luminosity of L$_{*}$ = 201 L$_{\odot}$ and
concluded that the star is probably a PPN.
The IRAS colours [12]\,-\,[25]\,=\,$+$2.45 and [25]\,-\,[60]\,=\, $-$0.34 indicate
that the star
falls into the region defined by \cite{GarciaLarioEtAl1997}.
The PAGB nature of IRAS\,18025\,-\,3906 with its optical counterpart has been
confirmed
by \cite{SuarezEtAl2006} using low-resolution spectra and is classified as a G1\,I
star.
\cite{VickersEtAl2015} found a distance of 5.20$\pm$0.70 kpc from an adopted luminosity of
6000$\pm$1500
L$_{\odot}$ respectively, locating this object probably at intermediate thin disk
population. The star has been monitored by the ASAS-3, the light curve includes 490
observations and 
does not show signs of a periodic variation.

\begin{figure}
\begin{minipage}{80mm}
\begin{center}
  \includegraphics[width=\columnwidth]{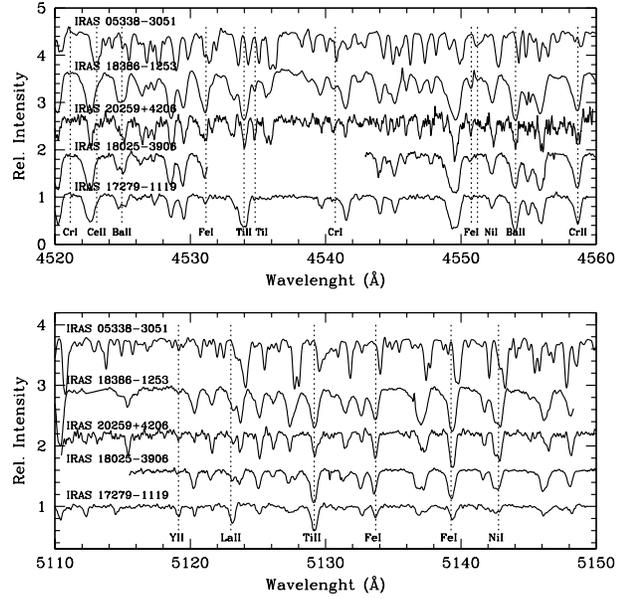}
  \caption{Representative spectra of the sample stars IRAS\,05338\,-\,3051,
IRAS\,17279\,-\,1119, IRAS\,18025\,-\,3906, IRAS\,18386\,-\,1253 and
IRAS\,20259\,+\,4206
of two different regions. 
The location of lines of certain important elements have been indicated by dashed
lines.
The stars are arranged in increasing order the effective temperature (top to bottom).}
  \label{fig:figure2}
\end{center}
\end{minipage}
\end{figure}

IRAS\,18386\,-\,1253 (GLPM 818) is a relatively unexplored object. Its IRAS
colours [12]\,-\,[25]\,=\, $+$2.13 and [25]\,-\,[60]\,=\, $+$0.04 place the
object into the PAGBs--PNs region in Fig. \ref{fig:figure1}. This
object has also been included in the Toru\'n Catalogue of Galactic PAGB \citep []{SzczerbaEtAl2007} as a
likely PAGBs. \cite{PereiraMiranda2007} classified the star as a PAGB by
comparing its
spectrum with that of the known PAGB GLMP 1078 and they assigned its spectral type as F5\,I.
\cite{VickersEtAl2015} found a distance of 6.83$\pm$1.03 kpc from an adopted luminosity of
6000$\pm$1500 L$_{\odot}$. Likewise IRAS\,18025\,-\,3906, IRAS 18386 - 1253 is classified as a possible
intermediate thin disk
population object. 
It has also been monitored for a long time by ASAS-3; its light curve includes some
539 observations. It does not display a periodic variation although it may have a
characteristic time of variation of about a hundred days.

IRAS\,20259\,+\,4206 (GLPM 998) was studied by \cite{SuarezEtAl2006} using
low-resolution
spectra and classified it as a F3\,I PAGB star. It was
also recorded as a likely PAGB in the Toru\'n Catalogue \citep []{SzczerbaEtAl2007}.
A failed search of H$_{2}$O and SiO masers in its envelope is reported by \cite{SuarezEtAl2007} and \cite{YoonEtAl2014}. \cite{VickersEtAl2015} considers this object as of the intermediate thin disk 
population.

It is clear that a detailed abundance analysis based on high resolution spectra 
for all these objects is key in order to confirm their evolutionary status.

In Table~\ref{tab:table1} we  list the equatorial and galactic coordinates, the $V$
magnitude, and IRAS fluxes for the stars in our sample. 

\begin{table*}
 \centering
  \caption{Basic data of the program stars. Numbers in parenthesis are the uncertainties in the flux.}
 \label{tab:table1}
\scriptsize{\begin{tabular}{lcccrcrrrr}
  \hline
  \hline
\multicolumn{1}{c}{No. IRAS}&
\multicolumn{1}{c}{$\alpha_{2000}$}&
\multicolumn{1}{c}{$\delta_{2000}$}&
\multicolumn{1}{c}{{\it V}}&
\multicolumn{1}{c}{{\it l}}&
\multicolumn{1}{c}{{\it b}}&
\multicolumn{1}{l}{F$_{12}$ $\mu$m}&
\multicolumn{1}{c}{F$_{25}$ $\mu$m}&
\multicolumn{1}{c}{F$_{60}$ $\mu$m}&
\multicolumn{1}{c}{F$_{100}$ $\mu$m}\\
\multicolumn{1}{c}{}&
\multicolumn{1}{c}{(h m s)}&
\multicolumn{1}{c}{(h m s)}&
\multicolumn{1}{c}{(mag)}&
\multicolumn{1}{c}{($^{0}$)}&
\multicolumn{1}{c}{($^{0}$)}&
\multicolumn{1}{c}{(Jy)}&
\multicolumn{1}{c}{(Jy)}&
\multicolumn{1}{c}{(Jy)}&
\multicolumn{1}{c}{(Jy)}\\
\hline
02528\,$+$\,4350&02 56 11.3&$+$44 02
52.1&10.68&145.41&$-$13.30&0.56(3)&2.38(3)&3.00(2)&2.22(1)\\
05338\,$-$\,3051&05 35 44.2&$-$30 49 35.3&9.30
&234.98&$-$28.78&0.35(3)&0.25(1)&0.40(1)&1.00(1)\\
17279\,$-$\,1119&17 30 46.9&$-$11 22 08.3&9.68
&013.23&$-$12.17&3.52(3)&2.90(3)&1.60(3)&1.98(1)\\
18025\,$-$\,3906&18 06 03.2&$-$39 05
56.6&11.73&353.26&$-$8.72&4.30(3)&41.20(3)&30.11(3)&7.68(3) \\
18386\,$-$\,1253&18 41 28.8&$-$12 50
52.0&13.10&20.36&$-$3.65&1.43(3)&10.19(3)&10.62(3)&48.27(1) \\
20259\,$+$\,4206&20 27 41.9&$+$42 16
42.0&13.75&80.39&$+$2.19&2.49(3)&8.82(3)&7.01(3)&135.40(1)\\
\hline
\end{tabular}}
\end{table*}

\section{Observations}
\label{sec:obser}

The spectra used in the present work were obtained at different observatories with a variety of high resolution spectrographs: 
The Fiber-fed Extended Range Optical Spectrograph (FEROS) mounted at the 2.2m
telescope at La Silla Observatory (Chile); 
the Fibre-fed Echelle Spectrograph (FIES) mounted on the 2.5m Nordic Optical
Telescope (NOT) at the Roque de los Muchachos Observatory at La Palma (Spain), 
the Tull echelle spectrograph \citep [] {TullEtAl1995} on the 2.7m telescope at the
McDonald Observatory (USA), and the Hanle Echelle
Spectrometer (HESP) mounted on the 2m optical-infrared Himalayan
Chandra Telescope (HCT) at the Indian Astronomical Observatory (IAO)
(India). 

IRAS\,02528\,+\,4350 and IRAS\,17279\,-\,1119 were observed on the nights of April 26,
and October 20, 2016. Two exposures were obtained for each object
to improve the S/N ratio. These spectra obtained with HESP cover the
3660-10550\AA\@ range and have a resolution power of
R($\lambda$/$\Delta$$\lambda$)\,$\sim$\,60,000.

IRAS\,05338\,-\,3051 was observed on February 10, 2017 with the Tull spectrograph of
McDonald
Observatory. Three exposures were combined to increase the S/N ratio. They
cover a range of wavelength of 3650--10350\AA\@ and a resolution of $\sim$60,000.

IRAS\,18025\,-\,3906 and IRAS\,18386\,-\,1253 were observed on September 30, 1998
and July 29, 2009 respectively. Their spectra were obtained with the Fiber-fed
Extended Range
Optical Spectrograph (FEROS) and cover a range of wavelength of 3900--9200 \AA\@ with
a
resolution of R$\sim$48,000. These spectra form part of the same collection of objects
studied 
by \cite{MolinaEtAl2014}.

IRAS\,20259\,+\,4206 was observed on June 24, 2014. Five spectra 
were obtained with an exposure time of 1800 secs each. Individual spectra have a S/N
$\sim$15. The combined spectrum has a S/N $\sim$30. The wavelength range obtained
with FIES is 3650--7150\AA\@ and it is distributed in 78 orders
with a resolution R$\sim$25,000. 
The spectra were reduced following the standard procedure and summed using the task
COMBINE=
MEDIAN to remove the cosmic ray hits. Because of the low signal-to-noise only a
limited number of line strengths for each identified elements were measured.

As a comparison, two 40\,\AA\@ regions, are displayed in Fig.~\ref{fig:figure2} for
each
of the spectra of the sample stars. 
The vertical dotted lines represent the positions at rest of some absorption lines of elements
such as 
\ion{Y}{ii}, \ion{La}{ii}, \ion{Ti}{ii}, \ion{Fe}{i}, \ion{Ba}{ii}, \ion{Cr}{ii},
\ion{Cr}{i} 
and \ion{Ni}{i} present in these spectral regions. In the two spectral regions shown 
in Fig.~\ref{fig:figure2}, the star IRAS\,05338\,-\,3051 is not present because its spectral 
features are very weak and barely detectable.

\section{ATMOSPHERIC PARAMETERS}
\label{sec:param}

The shape and strength of spectral lines depend strongly on the physical conditions
prevailing in the atmosphere of the star. Hence, good estimates of 
effective temperature, $T_{\rm eff}$, surface gravity, log\,$g$, and
microturbulence velocity, $\xi_{t}$ are necessary to derive accurate abundances. 
First estimates of these atmospheric parameters can be made from different approaches.

For the effective temperature a first approach can be obtained
for example from the spectral type and the calibrations of
\cite{SchmidtK1982}. 
Spectral types can be generally be found in the SIMBAD database
and/or form the work 
of \cite{SuarezEtAl2006}. Our calculations of $T_{\rm eff}$ are
9730K ($\sim$A0)
for
IRAS\,02528\,+\,4350, 4850K ($\sim$G5) 
for IRAS\,05338\,-\,3051, 7350K ($\sim$F2/3) for IRAS\,17279\,-\,1119, 5200K
($\sim$G2) for 
IRAS\,18025\,-\,3906, 6900K ($\sim$F5) for IRAS\,18386\,-\,1253 and 7200K ($\sim$F3)
for 
IRAS\,20259\,+\,4206.

A second approach to the effective temperature calculation is from the
spectroscopy calibrations of \cite{Molina2018} for PAGB stars. These calibrations 
enable to estimate the $T_{\rm eff}$, from the equivalent widths of
specific
absorption lines. For stars of intermediate temperature, we employed the
equivalent width of Ca\,IIK at 3933\,\AA, while for cooler stars the G-band at 4302\,\AA
~and
\ion{Ca}{i} at 4226\,\AA  ~are used.
From the sample studied, $T_{\rm eff}$ could be calculated by this approach for IRAS\,02528\,+\,4350
and IRAS\,
18025\,-\,3906, obtaining the values 7981$\pm$91K and 6196$\pm$91K respectively. The
remaining
objects do not have the above features in the observed spectral range.
 
A third independent approach to estimate $T_{\rm eff}$, and log\,$g$, is from the
Balmer lines 
profile fitting \citep []{Venn1995}. Theoretical Balmer profiles have been synthesized
using the {\tt
SPECTRUM}\footnote{http://www.appstate.edu/$\sim$grayro/spectrum/spectrum.html} 
code and a grid of {\tt ODFNEW} model atmospheres \citep []{CastelliKurucz2003}, and were used
to fit the observed profiles of H$\beta$ and H$\delta$. 
The solution is not unique as different pairs of $T_{\rm eff}$ -- log\,$g$ may produce
a good 
line fit. A locus of good solutions can be seen on the $T_{\rm eff}$--log\,$g$ plane. 
As underlying core emission may be present in some stars, we chose to fit only the
wings 
of the Balmer lines.
Additional loci are obtained by searching for $T_{\rm eff}$--log\,$g$ pairs that yield
consistent abundance for neutral and ionized lines of \ion{Mg}{i}/\ion{Mg}{ii}, 
\ion{Si}{i}/\ion{Si}{ii}, \ion{Sc}{i}/\ion{Sc}{ii},
\ion{Ti}{i}/\ion{Ti}{ii}, and \ion{Cr}{i}/\ion{Cr}{ii}.

\begin{figure}
\begin{center}
  \includegraphics[width=8.1cm,height=6.cm]{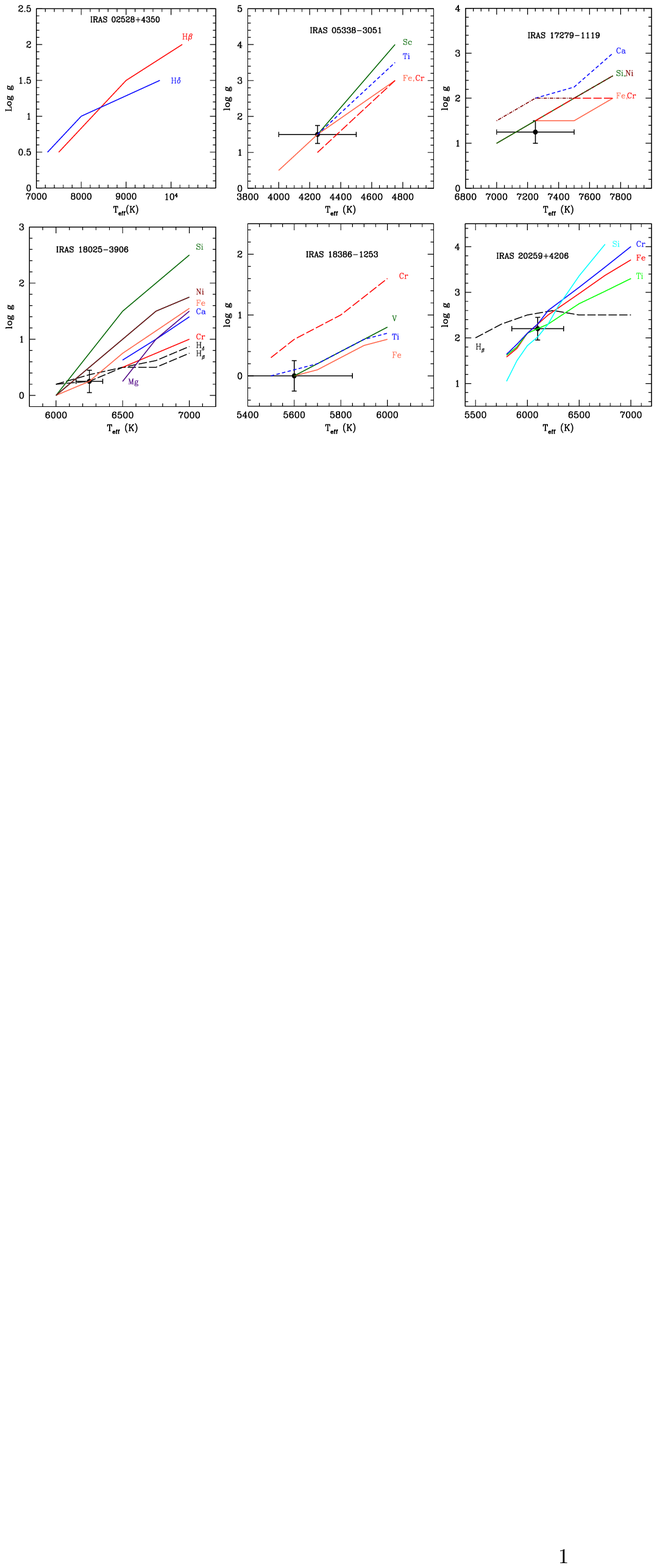}
  \caption{Locus of different solutions are represented on the $T_{\rm eff}$--log\,$g$
plane for each star of the sample. 
The filled circle with error bars indicate the adopted initial values of 
$T_{\rm eff}$ and log\,$g$ for the subsequent calculation of abundances. The colour version of this figure can be
seen online.}
  \label{fig:figure3}
\end{center}
\end{figure}

The intersection of these loci gives good estimates of the temperature and gravity.
For some stars however, the emission is so conspicuous that this criterion can not be used. 
We have shown in Fig.~\ref{fig:figure3} the loci of temperatures and gravities derived using
different species mentioned above. The adopted values for the temperature and gravity are indicated by the filled circle.   
With this method we estimated the initial value ranges of temperature and gravity for IRAS\,18025\,-\,3906 
(6125--6335\,K, 0.1--0.4) and for IRAS\,20259\,+\,4206 (5820--6380\,K, 1.9--2.4).
 IRAS\,02528\,+\,4350 a single solution (8441\,K; 1.13) is obtained from the intersection of H$\beta$ and H$\delta$
loci. The differences between the values adopted (7900\,K; 2.40) and the estimated from the intersection between the loci of the Balmer lines might be due to the poor quality of the observed spectrum.
For the remaining objects, the Balmer profiles are polluted by emission and could also affect the estimates of both parameters. The values of temperature and gravity were obtained by the intersection of loci of different species with two states of ionization. For IRAS\,05338\,-\,3051 (4000--4500\,K; 1.22--1.68) were estimated from the intersection between the Ti, Sc and Fe loci, for IRAS\,17279\,-\,1119 (7000--7500\,K; 1.0--1.4) from Si, Fe and Cr loci, and for IRAS\,18386\,-\,1253 (5400--5850\,K; 0.0--0.2) from Ti, V and Fe loci respectively.
These initial estimates of $T_{\rm eff}$ and log\,$g$ shall be refined by a detailed spectral analysis.

For the detailed abundance analysis, we employed the 2015 version of the spectral
code MOOG \citep []{Sneden1973} along with the grids of LTE plane-parallel {\tt ODFNEW}
model atmospheres. 

\begin{table*}
\begin{center}
  \caption{Adopted atmospheric parameters for all stars.}
 \label{tab:table2}
\begin{tabular}{lccccccr}
  \hline
  \hline
\multicolumn{1}{c}{No. IRAS}&
\multicolumn{1}{c}{$T_{\rm eff}$}&
\multicolumn{1}{c}{log~$g$}&
\multicolumn{1}{c}{$\xi_{t}$}&
\multicolumn{1}{c}{[Fe/H]}&
\multicolumn{1}{c}{\it {V$_{r}$(hel)}}&
\multicolumn{1}{c}{\it {V$_{r}$(lsr)}}&
\multicolumn{1}{c}{Date}\\
\multicolumn{1}{c}{}&
\multicolumn{1}{c}{(K)}&
\multicolumn{1}{c}{}&
\multicolumn{1}{c}{(km s$^{-1}$)}&
\multicolumn{1}{c}{}&
\multicolumn{1}{c}{(km s$^{-1}$)}&
\multicolumn{1}{c}{(km s$^{-1}$)}&
\multicolumn{1}{c}{}\\
\hline
02528\,$+$\,4350 & 7900 & 2.40 & 2.00 & $-$0.90 & $-$0.9 & $-$2.4  & October 20,
2016\\
05338\,$-$\,3051 & 4250 & 1.50 & 2.50 & $-$1.32 & $+$5.2 & $-$14.7 & February 10,
2017\\
17279\,$-$\,1119 & 7250 & 1.25 & 4.40 & $-$0.59 & $+$61.6 & $+$76.4 & April 26, 2016
\\
18025\,$-$\,3906 & 6250 & 0.25 & 4.00 & $-$0.50 & $-$120.2 & $-$113.0 & September 30,
1998 \\
18386\,$-$\,1253 & 5600 & 0.00 & 4.30 & $-$0.18 & $+$83.2  & $+$97.6  & July 29, 2009
\\
20259\,$+$\,4206 & 6100 & 2.20 & 2.36 & $-$0.17 & $-$15.6  & $+$1.4   & June 24, 2014
\\
\hline
\end{tabular}
\end{center}
\end{table*}
                    
\begin{figure}
\begin{center}
  \includegraphics[width=8.cm,height=8cm]{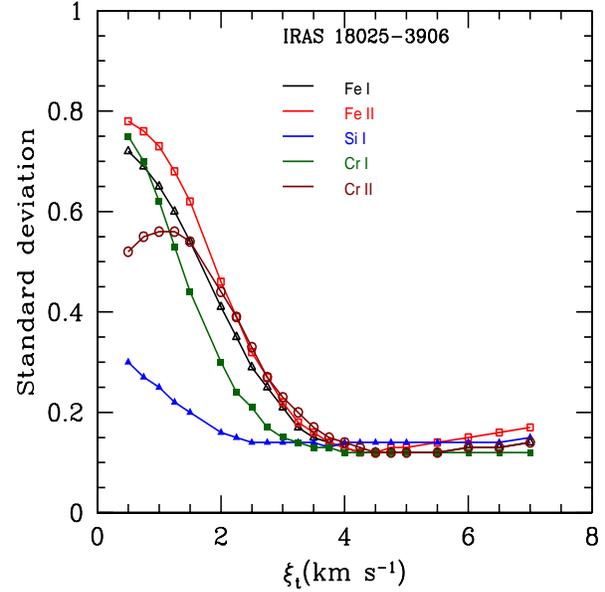}
  \caption{Determination of microturbulence velocity for the star IRAS\,
18025\,-\,3906 using the plot of standard deviation as a function of
microturbulence velocity $\xi_{t}$ for several species  in the 0 to 8 kms$^{-1}$
range.
The colour figure can be viewed in the online version.}
  \label{fig:figure4}
\end{center}
\end{figure}

Temperature and gravity can be estimated from accurate measurements of equivalent
widths
(EWs) for a set of Fe lines with well determined atomic data (log\,$gf$ and $\chi$
(eV)). 
We restricted to EWs with values between 10 to 200 m\AA\@ to avoid NLTE effects on strong lines. 
By eliminating the dependence of calculated abundances
on lower excitation potentials and requiring that the neutral and ionized lines give
the same abundances, i.e. $\log$\,n(\ion{Fe}{i}) $=$ $\log$\,n(\ion{Fe}{ii}) one can get
good estimates of $T_{\rm eff}$, log\,$g$ respectively.
In order to confirm the value of gravity, the equilibrium condition can be extended to
other elements 
with two ionization states (such as Mg, Si, Ca, Ti, Cr, Ni). The microturbulence
velocity was estimated by requiring that the derived abundances are independent of
line strengths. 
In addition we can confirm the microturbulence velocity value by plotting $\sigma_{X}$
vs. $\xi_{t}$ and 
following the method described in \cite{SahinLambert2009}. Standard deviation for
elements like Fe, Si
and Cr leads to a minimum value of microturbulence velocity in the 0 to 8 kms$^{-1}$
range. 
Fig. \ref{fig:figure4} illustrates the method for the star IRAS\,18025\,-\,3906\@, in
which a minimum error for $\xi_{t}$ of $\sim$4.0 kms$^{-1}$ is found.

An error in the EWs of about 8 to 10\%, corresponds to an uncertainty in the effective
temperature of 
$\pm$250K, to $\pm$0.25\@ in the superficial gravity and to $\pm$0.5\@ kms$^{-1}$ in
the 
microturbulence velocity of $\pm$0.5\@ kms$^{-1}$. Our final adopted values of the
atmospheric parameters for the 
stars in our sample are listed in Table~\ref{tab:table2}.

\section{DETERMINATION OF ATMOSPHERIC ABUNDANCES}
\label{sec:abund}

In the determination of atmospheric chemical abundances we use the equivalent widths of 141, 152,
205, 178 and 84 absorption lines identified in the spectra of IRAS\,05338\,-\,3051,
IRAS\,17279\,-\,1119,
IRAS\,18025\,-\,3906, IRAS\,18386\,-\,1253 and IRAS\,20259\,+\,4206 respectively.
For this purpose we use the task {\tt ABFIND} of {\tt MOOG} code \citep []{Sneden1973}.
Some lines are affected by hyperfine structure (HFs) e.g. Sc, Mn, Cu, Eu,
therefore in these cases, the corresponding spectral regions have been synthesized using the  {\tt
SPECTRUM} code.
For IRAS\,02528\,+\,4350 the lines were so broad and the spectrum poor that the
chemical abundances were determined exclusively from spectral synthesis.  

The chemical abundances derived for the program stars are presented in
Tables~\ref{tab:table3} and \ref{tab:table4}.

These tables identify the chemical species, the solar abundance from
 \cite {AsplundEtAl2009}, the ratio of the abundances relative to hydrogen,
the total uncertainty in the determination of abundances from the atmospheric model,
the
number of lines used or synthesized for each element, and abundances relative to iron.

\begin{table*}
\caption{Elemental abundances for IRAS\,02528\,+\,4350, IRAS\,05338\,-\,3051 and
IRAS\,17279\,+\,1119.}
\label{tab:table3}
\begin{center}
\scriptsize{\begin{tabular}{lcrrrrrrrrr}
\noalign{\smallskip}
\hline
\noalign{\smallskip}
\noalign{\smallskip}
 &
&\multicolumn{3}{c}{IRAS\,02528\,+\,4350}&\multicolumn{3}{c}{IRAS\,05338\,-\,3051}
&\multicolumn{3}{c}{IRAS\,17279\,-\,1119}\\
\cline{3-5} \cline {6-8} \cline {9-11} \\
\multicolumn{1}{l}{Species}&
\multicolumn{1}{c}{$\log \epsilon_{\odot}$}&
\multicolumn{1}{r}{[X/H]}&
\multicolumn{1}{c}{N}&
\multicolumn{1}{r}{[X/Fe]}&
\multicolumn{1}{c}{[X/H]}&
\multicolumn{1}{r}{N}&
\multicolumn{1}{r}{[X/Fe]}&
\multicolumn{1}{r}{[X/H]}&
\multicolumn{1}{c}{N}&
\multicolumn{1}{r}{[X/Fe]}\\
           \noalign{\smallskip}
            \hline
            \noalign{\smallskip}
\ion{C}{i}     & 8.43 & $-0.13$$\pm$0.10 & syn(2) &$+0.77$&                    &   
    &          
& $-0.18$$\pm$0.12 & 7 &$-0.41$\\
C (C$_{2}$/CH) & 8.43 &                    &        &        & $-1.13$$\pm$0.10 &
syn(4) & $+0.19$         
&                    &   &        \\
\ion{N}{i}      & 7.83 &                    &        &        &                    &  
     &          
& $+0.17$$\pm$0.20 & 2 &$+0.76$\\
N (CN)         & 7.83 &                    &        &        & $-0.63$$\pm$0.10 &
syn(1) & $+0.69$         
&                    &   &        \\
\ion{O}{i}         & 8.69 & $-0.09$$\pm$0.20 & syn(3) &$+0.81$& $-0.48$$\pm$0.06
&  3     & $+0.84$ 
& $-0.36$$\pm$0.03 & 3 &$+0.23$\\
\ion{Na}{i}         & 6.24 & $-0.11$$\pm$0.20 & syn(1) &$+0.79$& $-1.47$$\pm$0.07
&  4     & $-0.15$ 
& $+0.03$$\pm$0.06 & 3 &$+0.56$\\
\ion{Mg}{i}        & 7.60 & $-0.82$$\pm$0.20 & syn(2) &$+0.08$& $-0.90$$\pm$0.10
&  4     & $+0.37$ 
& $-0.41$$\pm$0.11 & 2 &$+0.18$\\
\ion{Mg}{ii}       & 7.60 & $-0.82$$\pm$0.20 & syn(1) &$+0.08$&                   
&        &                  
&                    &   &        \\
\ion{Al}{i}       & 6.45 &                    &        &        & $-1.53$$\pm$0.03 &
 2     & $-0.21$ 
&                    &   &        \\
\ion{Si}{i}       & 7.51 & $-0.71$$\pm$0.20 & syn(2) &$+0.19$& $-0.72$$\pm$0.04 &
 2     & $+0.60$ 
& $-0.16$$\pm$0.03 & 1 &$+0.43$\\
\ion{Si}{ii}      & 7.51 &                    &        &        &                    &
       &          
& $-0.12$$\pm$0.22 & 2 &$+0.47$\\
\ion{S}{i}        & 7.12 &                    &        &        &                    &
       &          
& $-0.02$$\pm$0.00 & 1 &$+0.57$\\
\ion{Ca}{i}       & 6.34 & $-0.58$$\pm$0.20 & syn(1) &$+0.32$& $-1.19$$\pm$0.11 &
 12    & $+0.13$ 
& $-0.45$$\pm$0.08 & 7 &$+0.13$\\
\ion{Ca}{ii}      & 6.34 &                    &        &        &                    &
       &          
& $-0.51$$\pm$0.04 & 1 &$+0.08$\\
\ion{Sc}{i}       & 3.15 &                    &        &        & $-1.29$$\pm$0.08 &
 1     & $+0.00$         
&                    &   &        \\
\ion{Sc}{ii}       & 3.15 &                    &        &        & $-1.21$$\pm$0.08
&  5     & $+0.11$ 
& $-0.35$$\pm$0.10 & 9 &$+0.24$\\
\ion{Ti}{i}     & 4.95 &                    &        &        & $-1.06$$\pm$0.10 & 
9     & $+0.26$         
&                    &   &        \\
\ion{Ti}{ii}        & 4.95 & $-0.95$$\pm$0.20 & syn(2) &$-0.05$& $-1.12$$\pm$0.08
&  3     & $+0.20$ 
& $-0.11$$\pm$0.12 &13 &$+0.48$\\
\ion{V}{i}        & 3.93 &                    &        &        & $-0.86$$\pm$0.09 &
 8     & $+0.46$         
&                    &   &        \\
\ion{V}{ii}        & 3.93 &                    &        &        &                   
&        &          
& $-0.27$$\pm$0.08 & 1 &$+0.32$\\
\ion{Cr}{i}       & 5.64 & $-0.97$$\pm$0.20 & syn(1) &$-0.07$& $-1.15$$\pm$0.07 &
 5     & $+0.17$ 
& $-0.31$$\pm$0.04 & 1 &$+0.28$\\
\ion{Cr}{ii}       & 5.64 & $-0.97$$\pm$0.20 & syn(1) &$-0.07$& $-1.04$$\pm$0.10
&  1     & $+0.28$ 
& $-0.25$$\pm$0.10 &11 &$+0.33$\\
\ion{Mn}{i}        & 5.43 &                    &        &        & $-1.49$$\pm$0.09
&  2     & $-0.17$ 
& $-1.18$$\pm$0.10 & 3 &$-0.59$\\
\ion{Fe}{i}        & 7.50 & $-0.90$$\pm$0.20 & syn(4) &        & $-1.34$$\pm$0.09
&  37    &                  
& $-0.63$$\pm$0.20 & 36 &       \\
\ion{Fe}{ii}        & 7.50 & $-0.90$$\pm$0.20 & syn(4) &        & $-1.30$$\pm$0.08
&  6     &                  
& $-0.54$$\pm$0.14 & 20&        \\
\ion{Co}{i}       & 4.99 &                    &        &        & $-1.24$$\pm$0.09 &
 3     & $+0.08$         
&                    &   &        \\
\ion{Ni}{i}        & 6.22 &                    &        &        & $-0.97$$\pm$0.02
&  5     & $+0.35$ 
& $-0.15$$\pm$0.13 & 2 &$+0.44$\\
\ion{Ni}{ii}        & 6.22 &                   &        &        &                   
&        &          
& $-0.28$$\pm$0.04 & 1 &$+0.31$\\
\ion{Cu}{i}        & 4.19 &                    &        &        & $-1.48$$\pm$0.10
& syn(1) & $-0.16$         
&                    &   &        \\
\ion{Zn}{i}        & 4.56 &                    &        &        & $-1.59$$\pm$0.08
&  2     & $-0.27$ 
& $-0.10$$\pm$0.05 & 1 &$+0.49$\\
\ion{Sr}{ii}        & 2.87 & $-0.97$$\pm$0.20 & syn(2) &$-0.07$&                   
&        &                  
&                    &   &        \\
\ion{Y}{ii}        & 2.21 &                    &        &        & $-1.02$$\pm$0.05
&  3     & $+0.30$ 
& $+0.41$$\pm$0.09 & 6 &$+1.00$\\
\ion{Zr}{ii}        & 2.58 &                    &        &        & $-0.57$$\pm$0.02
&  3     & $+0.75$ 
& $+0.24$$\pm$0.12 & 2 &$+0.82$\\
\ion{Mo}{i}        & 1.88 &                    &        &        & $-0.77$$\pm$0.06
&  2     & $+0.55$         
&                    &   &        \\
\ion{Ba}{ii}        & 2.18 & $-1.08$$\pm$0.20 & syn(1) &$-0.18$& $-1.13$$\pm$0.07
&  1     & $+0.19$ 
& $+0.32$$\pm$0.20 & 1 &$+0.91$\\
\ion{La}{ii}        & 1.10 &                    &        &        & $-0.83$$\pm$0.02
&  2     & $+0.49$ 
& $+0.51$$\pm$0.17 & 2 &$+1.09$\\
\ion{Ce}{ii}        & 1.58 &                    &        &        & $-0.93$$\pm$0.09
&  3     & $+0.39$         
& $+0.38$$\pm$0.09 & 3 &$+0.97$\\
\ion{Pr}{ii}        & 0.72 &                    &        &        & $-0.57$$\pm$0.10
& syn(3) & $+0.75$         
& $+0.54$$\pm$0.12 & 5 &$+1.12$\\
\ion{Nd}{ii}        & 1.42 &                    &        &        & $-0.83$$\pm$0.04
&  8     & $+0.49$         
& $+0.40$$\pm$0.04 & 2 &$+0.99$\\
\ion{Sm}{ii}        & 0.96 &                    &        &        & $-0.55$$\pm$0.00
&  2     & $+0.77$         
& $+0.15$$\pm$0.09 & 2 &$+0.74$\\
\ion{Eu}{ii}        & 0.52 &                    &        &        & $-0.43$$\pm$0.12
&  2     & $+0.89$         
& $+0.17$$\pm$0.05 & 1 &$+0.76$\\
\ion{Gd}{ii}        & 1.07 &                    &        &        & $-0.89$$\pm$0.04
&  1     & $+0.53$         
&                    &   &        \\
            \noalign{\smallskip}
            \hline
            \noalign{\smallskip}
\end{tabular}}
\end{center}
\end{table*}

\begin{table*}
\caption{Elemental abundances for IRAS\,18025\,-\,3906, IRAS\,18386\,-\,1253 and
IRAS\,20259\,+\,4206.}
\label{tab:table4}
\begin{center}
\scriptsize{\begin{tabular}{lccrccrrccr}
\noalign{\smallskip}
\hline
\noalign{\smallskip}
\noalign{\smallskip}
 &
&\multicolumn{3}{c}{IRAS\,18025\,-\,3906}&\multicolumn{3}{c}{IRAS\,18386\,-\,1253}
&\multicolumn{3}{c}{IRAS\,20259\,+\,4206}\\
\cline{3-5} \cline {6-8} \cline {9-11}\\
\multicolumn{1}{l}{Species}&
\multicolumn{1}{c}{$\log \epsilon_{\odot}$}&
\multicolumn{1}{c}{[X/H]}&
\multicolumn{1}{r}{N}&
\multicolumn{1}{r}{[X/Fe]}&
\multicolumn{1}{c}{[X/H]}&
\multicolumn{1}{r}{N}&
\multicolumn{1}{r}{[X/Fe]}&
\multicolumn{1}{c}{[X/H]}&
\multicolumn{1}{c}{N}&
\multicolumn{1}{r}{[X/Fe]}\\
           \noalign{\smallskip}
            \hline
            \noalign{\smallskip}
\ion{C}{i} & 8.43 & $-0.04$$\pm$0.16 & 20   &$+0.46$ & $+0.20$$\pm$0.97 &   12
&$+0.39$& $-0.67$$\pm$0.03 &  2   &$-0.50$\\
\ion{N}{i} & 7.83 & $+0.23$$\pm$0.18 &  3   &$+0.74$ & $+0.66$$\pm$0.12 &    3
&$+0.84$&                    &      & \\
\ion{O}{i} & 8.69 & $+0.06$$\pm$0.10 &syn(3)&$+0.56$ & $-0.21$$\pm$0.01 &    2
&$-0.03$&                    &      & \\
\ion{Na}{i} & 6.24 & $-0.09$$\pm$0.04 &  4   &$+0.41$ & $+0.33$$\pm$0.04 &    1
&$+0.51$& $+0.03$$\pm$0.08 &  2   &$+0.20$\\
\ion{Mg}{i} & 7.60 & $-0.32$$\pm$0.09 &  3   &$+0.19$ & $+0.01$$\pm$0.11 &    2
&$+0.19$& $-0.36$$\pm$0.00 &  2   &$-0.19$\\
\ion{Mg}{ii}& 7.60 & $-0.06$$\pm$0.09 &  1   &$+0.44$ &                    &      &
       &                    &      & \\
\ion{Al}{i} & 6.45 &                    &      &         &                    &      &
       & $-0.38$$\pm$0.20 &  2   &$-0.22$\\
\ion{Si}{i} & 7.51 & $+0.01$$\pm$0.14 & 14   &$+0.52$ & $-0.00$$\pm$0.16 &   11
&$+0.18$& $+0.09$$\pm$0.14 &  4   &$+0.26$\\
\ion{Si}{ii}& 7.51 & $-0.17$$\pm$0.08 &  2   &$+0.33$ &                    &      &
       & $+0.21$$\pm$0.09 &  2   &$+0.37$\\
\ion{S}{i} & 7.12 & $+0.38$$\pm$0.09 &  6   &$+0.88$ & $+0.20$$\pm$0.10
&syn(3)&$+0.38$& $-0.10$$\pm$0.10 &syn(1)&$-0.07$\\
\ion{Ca}{i} & 6.34 & $-0.48$$\pm$0.14 & 13   &$+0.02$ & $-0.48$$\pm$0.10 &   10
&$-0.30$& $+0.02$$\pm$0.06 &  4   &$+0.18$\\
\ion{Ca}{ii}& 6.34 & $-0.45$$\pm$0.04 &  1   &$+0.05$ &                    &      &
       &                    &      & \\
\ion{Sc}{ii}& 3.15 & $-1.15$$\pm$0.10 &syn(1)&$-0.65$ & $-1.03$$\pm$0.03 &    2
&$-0.85$& $-0.79$$\pm$0.12 &  2   &$-0.62$\\
\ion{Ti}{i} & 4.95 &                    &      &         & $-0.73$$\pm$0.07 &    2
&$-0.55$& $-0.29$$\pm$0.15 &  4   &$-0.13$\\
\ion{Ti}{ii}& 4.95 & $-0.48$$\pm$0.10 &  6   &$+0.02$ & $-0.76$$\pm$0.06 &    1
&$-0.58$& $-0.28$$\pm$0.13 &  3   &$-0.11$\\
\ion{V}{i} & 3.93 &                    &      &         & $-0.59$$\pm$0.07 &    2
&$-0.41$& $+0.16$$\pm$0.02 &  1   &$+0.33$\\
\ion{V}{ii} & 3.93 & $-0.50$$\pm$0.07 &  2   &$-0.00$ & $-0.58$$\pm$0.10 &    2
&$-0.40$&                    &      & \\
\ion{Cr}{i} & 5.64 & $-0.60$$\pm$0.07 &  7   &$-0.09$ & $-0.46$$\pm$0.05 &    5
&$-0.28$& $-0.37$$\pm$0.10 &  4   &$-0.20$\\
\ion{Cr}{ii}& 5.64 & $-0.39$$\pm$0.14 &  9   &$+0.11$ & $-0.72$$\pm$0.15 &    2
&$-0.54$& $-0.39$$\pm$0.08 &  4   &$-0.23$\\
\ion{Mn}{i} & 5.43 & $-0.73$$\pm$0.09 &syn(1)&$-0.23$ & $-0.73$$\pm$0.15 &    3
&$-0.55$& $-0.67$$\pm$0.13 &  2   &$-0.51$\\
\ion{Fe}{i} & 7.50 & $-0.50$$\pm$0.12 & 56   &         & $-0.19$$\pm$0.11 &   61 &
       & $-0.17$$\pm$0.11 & 19   & \\
\ion{Fe}{ii}& 7.50 & $-0.51$$\pm$0.17 & 18   &         & $-0.17$$\pm$0.11 &   12 &
       & $-0.07$$\pm$0.06 &  5   & \\
\ion{Co}{i} & 4.99 &                    &      &         & $-0.60$$\pm$0.08 &    1
&$-0.42$& $+0.12$$\pm$0.04 &  2   &$+0.29$\\
\ion{Ni}{i} & 6.22 & $-0.43$$\pm$0.12 & 13   &$+0.08$ & $-0.33$$\pm$0.15 &   14
&$-0.15$& $-0.44$$\pm$0.08 &  6   &$-0.28$\\
\ion{Ni}{ii}& 6.22 & $-0.56$$\pm$0.04 &  1   &$-0.06$ &                    &      &
       &                    &      & \\
\ion{Cu}{i} & 4.19 & $-0.38$$\pm$0.04 &  1   &$+0.13$ & $-0.37$$\pm$0.20
&syn(1)&$-0.19$& $-0.33$$\pm$0.04 &  1   &$-0.16$\\
\ion{Zn}{i} & 4.56 & $-0.33$$\pm$0.12 &  4   &$+0.18$ & $-0.21$$\pm$0.19 &    2
&$-0.02$& $-0.11$$\pm$0.06 &  2   &$-0.06$\\
\ion{Y}{ii} & 2.21 & $-1.31$$\pm$0.16 &  3   &$-0.80$ & $-1.25$$\pm$0.11 &    3
&$-1.07$& $-0.64$$\pm$0.01 &  2   &$-0.48$\\
\ion{Zr}{ii}& 2.58 & $-1.35$$\pm$0.04 &  1   &$-0.84$ & $-0.59$$\pm$0.04 &    1
&$-0.41$& $-1.32$$\pm$0.01 &  1   &$-1.15$\\
\ion{Ba}{ii}& 2.18 & $-0.52$$\pm$0.07 &  1   &$-0.01$ &                    &      &
       & $-0.71$$\pm$0.15 &  3   &$-0.54$\\
\ion{La}{ii}& 1.10 &                    &      &         & $-0.88$$\pm$0.05 &    1
&$-0.70$& $+0.01$$\pm$0.05 &  1   &$+0.18$\\
\ion{Ce}{ii}& 1.58 & $-0.92$$\pm$0.07 &  1   &$-0.41$ & $-1.09$$\pm$0.09 &    4
&$-0.91$&                    &      & \\
\ion{Nd}{ii}& 1.42 & $-0.42$$\pm$0.19 &  2   &$+0.09$ & $-1.24$$\pm$0.04 &    1
&$-1.06$&                    &      & \\
\ion{Sm}{ii}& 0.96 & $-0.63$$\pm$0.06 &  1   &$-0.12$ & $-1.22$$\pm$0.10 &    2
&$-1.04$& $-0.23$$\pm$0.06 &  1   &$-0.06$\\
\ion{Eu}{ii}& 0.52 & $-0.33$$\pm$0.06 &  1   &$+0.18$ & $-0.33$$\pm$0.10
&syn(1)&$-0.15$&                    &      & \\
            \noalign{\smallskip}
            \hline
            \noalign{\smallskip}
\end{tabular}}
\end{center}
\end{table*}

\begin{table*}
\begin{minipage}{170mm}
\caption{Sensitivity of abundances to the uncertainties in the model parameters for
a range of temperature covering our sample stars.}
\label{tab:table5}
\begin{center}
\scriptsize{\begin{tabular}{lcccccccccccc}
\noalign{\smallskip}
\hline
\noalign{\smallskip}
\noalign{\smallskip}
 
&\multicolumn{4}{c}{IRAS\,05338\,-\,3051}&\multicolumn{4}{c}{IRAS\,18386\,-\,1253}
&\multicolumn{4}{c}{IRAS\,18025\,-\,3906}\\
  &  &  & (4250K)  & &  &  & (5600K) & &  &  & (6250K) &  \\
\multicolumn{1}{l}{Species}&
\multicolumn{1}{c}{$\Delta T_{\rm eff} $}&
\multicolumn{1}{c}{$\Delta \log $g$ $}&
\multicolumn{1}{c}{$\Delta \xi_{t} $}&
\multicolumn{1}{c}{$\sigma_{tot}$}&
\multicolumn{1}{c}{$\Delta T_{\rm eff} $}&
\multicolumn{1}{c}{$\Delta \log $g$ $}&
\multicolumn{1}{c}{$\Delta \xi_{t} $}&
\multicolumn{1}{c}{$\sigma_{tot}$}&
\multicolumn{1}{c}{$\Delta T_{\rm eff}$ }&
\multicolumn{1}{c}{$\Delta \log $g$ $}&
\multicolumn{1}{c}{$\Delta \xi_{t} $}&
\multicolumn{1}{c}{$\sigma_{tot}$}\\
\multicolumn{1}{c}{}&
\multicolumn{1}{c}{+250}&
\multicolumn{1}{c}{+0.25}&
\multicolumn{1}{c}{+0.50}&
\multicolumn{1}{c}{}&
\multicolumn{1}{r}{+250}&
\multicolumn{1}{r}{+0.25}&
\multicolumn{1}{c}{+0.50}&
\multicolumn{1}{c}{}&
\multicolumn{1}{c}{+250}&
\multicolumn{1}{c}{+0.25}&
\multicolumn{1}{r}{+0.50}&
\multicolumn{1}{c}{}\\
           \noalign{\smallskip}
            \hline
            \noalign{\smallskip}
\ion{C}{i} &         &         &         &       &$+0.07$ &$-0.06$ &$+0.03$
&$0.09$&$-0.04$ &$-0.02$ &$+0.03$&$0.05$ \\
\ion{N}{i} &         &         &         &       &$+0.18$ &$-0.08$ &$+0.05$
&$0.20$&$+0.10$ &$-0.05$ &$+0.03$&$0.12$ \\
\ion{O}{i} &$-0.12$ &$+0.10$ &$+0.02$ &$0.16$&$-0.09$ &$-0.06$ &$+0.01$
&$0.10$&$+0.02$ &$-0.05$ &$+0.03$&$0.06$ \\
\ion{Na}{i} &$-0.21$ &$-0.04$ &$+0.05$&$0.22$&$-0.12$ &$+0.03$ &$+0.02$
&$0.13$&$-0.17$ &$+0.05$ &$+0.02$&$0.18$ \\
\ion{Mg}{i} &$-0.16$ &$-0.02$ &$+0.06$&$0.17$&$-0.09$ &$+0.02$ &$+0.07$
&$0.12$&$-0.15$ &$+0.04$ &$+0.04$&$0.16$ \\
\ion{Mg}{ii}&         &         &        &       &         &         &         &      
&$+0.08$ &$-0.05$ &$+0.07$&$0.12$ \\
\ion{Si}{i} &$+0.01$ &$+0.07$ &$+0.01$&$0.07$&$-0.11$ &$+0.03$ &$+0.02$
&$0.11$&$-0.17$ &$+0.05$ &$+0.02$&$0.17$ \\
\ion{Si}{ii}&         &         &        &       &         &         &         &      
&$+0.09$ &$-0.05$ &$+0.02$&$0.11$ \\
\ion{S}{i} &         &         &         &       &$+0.01$ &$-0.04$ &$+0.07$
&$0.08$&$-0.11$ &$+0.01$ &$+0.03$&$0.12$ \\
\ion{Ca}{i} &$-0.28$ &$-0.04$ &$+0.18$&$0.31$&$-0.15$ &$+0.03$ &$+0.04$
&$0.16$&$-0.23$ &$+0.05$ &$+0.04$&$0.24$ \\
\ion{Ca}{ii}&         &         &        &       &         &         &         &      
&$-0.02$ &$-0.06$ &$+0.06$&$0.08$ \\
\ion{Sc}{ii}&$-0.25$ &$-0.00$ &$+0.01$&$0.25$&$-0.07$ &$-0.07$ &$+0.07$
&$0.12$&$-0.14$ &$-0.06$ &$+0.06$&$0.16$ \\
\ion{Ti}{i} &$-0.30$ &$-0.04$ &$+0.08$&$0.31$&$-0.22$ &$+0.04$ &$+0.01$ &      
&         &         &        &        \\
\ion{Ti}{ii}&$-0.06$ &$-0.17$ &$+0.17$&$0.25$&$-0.08$ &$-0.06$ &$+0.09$
&$0.13$&$-0.12$ &$-0.06$ &$+0.08$&$0.16$ \\
\ion{V}{i} &$-0.30$ &$-0.00$ &$+0.08$ &$0.31$&$-0.23$ &$+0.04$ &$+0.01$ &      
&         &         &        &        \\
\ion{V}{ii} &         &         &        &       &$-0.05$ &$-0.06$ &$+0.01$
&$0.08$&$-0.11$ &$-0.06$ &$+0.03$&$0.13$ \\
\ion{Cr}{i} &$-0.23$ &$-0.03$ &$+0.07$&$0.24$&$-0.20$ &$+0.04$ &$+0.05$
&$0.21$&$-0.24$ &$+0.04$ &$+0.02$&$0.25$ \\
\ion{Cr}{ii}&$+0.10$ &$+0.18$ &$+0.02$&$0.21$&$-0.06$ &$-0.06$ &$+0.06$
&$0.10$&$-0.06$ &$-0.06$ &$+0.06$&$0.10$ \\
\ion{Mn}{i} &$-0.29$ &$-0.01$ &$+0.12$&$0.32$&$-0.17$ &$-0.00$ &$+0.01$
&$0.17$&$-0.22$ &$+0.04$ &$+0.01$&$0.23$ \\
\ion{Fe}{i} &$-0.22$ &$+0.01$ &$+0.10$&$0.24$&$-0.16$ &$+0.03$ &$+0.05$
&$0.17$&$-0.22$ &$+0.04$ &$+0.03$&$0.22$ \\
\ion{Fe}{ii}&$+0.13$ &$+0.13$ &$+0.07$&$0.19$&$-0.01$ &$-0.07$ &$+0.08$
&$0.11$&$-0.07$ &$-0.06$ &$+0.05$&$0.10$ \\
\ion{Co}{i} &$-0.20$ &$+0.05$ &$+0.02$&$0.21$&$-0.17$ &$+0.04$ &$+0.01$ &      
&         &         &        &        \\
\ion{Ni}{i} &$-0.25$ &$+0.07$ &$+0.17$&$0.31$&$-0.16$ &$+0.03$ &$+0.07$
&$0.17$&$-0.22$ &$+0.04$ &$+0.01$&$0.22$ \\
\ion{Ni}{ii}&         &         &        &       &         &         &         &      
&$-0.06$ &$-0.06$ &$+0.03$&$0.09$ \\
\ion{Cu}{i} &         &         &        &       &$-0.20$ &$+0.04$ &$+0.04$
&$0.21$&$-0.26$ &$+0.05$ &$+0.02$&$0.27$ \\
\ion{Zn}{i} &$+0.05$ &$+0.12$ &$+0.06$&$0.15$&$-0.16$ &$+0.03$ &$+0.05$
&$0.17$&$-0.22$ &$+0.04$ &$+0.02$&$0.22$ \\
\ion{Y}{ii} &$-0.09$ &$+0.17$ &$+0.14$&$0.24$&$-0.09$ &$-0.06$ &$+0.03$
&$0.11$&$-0.16$ &$-0.05$ &$+0.01$&$0.17$ \\
\ion{Zr}{i} &$-0.28$ &$+0.01$ &$+0.04$&$0.29$&         &         &         &      
&         &         &        &        \\
\ion{Zr}{ii}&         &         &        &       &$-0.10$ &$-0.06$ &$+0.13$
&$0.17$&$-0.13$ &$-0.06$ &$+0.00$&$0.14$ \\
\ion{Mo}{i} &$-0.21$ &$-0.02$ &$+0.01$&$0.21$&         &         &         &      
&         &         &        &        \\
\ion{Ba}{ii}&$-0.21$ &$+0.18$ &$+0.16$&$0.32$&         &         &         &      
&$-0.21$ &$+0.01$ &$+0.12$&$0.24$ \\
\ion{La}{ii}&$-0.16$ &$+0.10$ &$+0.03$&$0.19$&$-0.13$ &$-0.06$ &$+0.09$ &      
&         &         &        &        \\
\ion{Ce}{ii}&$-0.11$ &$+0.19$ &$+0.02$&$0.22$&$-0.13$ &$-0.05$ &$+0.02$
&$0.14$&$-0.22$ &$-0.04$ &$+0.01$&$0.22$ \\
\ion{Nd}{ii}&$-0.15$ &$+0.18$ &$+0.06$&$0.25$&$-0.15$ &$-0.05$ &$+0.04$
&$0.16$&$-0.27$ &$-0.02$ &$+0.01$&$0.27$ \\
\ion{Sm}{ii}&$-0.15$ &$+0.17$ &$+0.07$&$0.24$&$-0.13$ &$-0.05$ &$+0.01$
&$0.14$&$-0.24$ &$-0.04$ &$+0.00$&$0.24$ \\
\ion{Eu}{ii}&$-0.08$ &$-0.02$ &$+0.03$&$0.09$&$-0.10$ &$-0.06$ &$+0.10$
&$0.16$&$-0.20$ &$-0.04$ &$-0.19$&$0.28$ \\
\ion{Gd}{ii}&$-0.13$ &$+0.12$ &$+0.03$&       &         &         &         &      
&         &         &        &        \\
            \noalign{\smallskip}
            \hline
            \noalign{\smallskip}
\end{tabular}}
\end{center}
\end{minipage}
\end{table*}

The abundances of the elements in Tables~\ref{tab:table3} and \ref{tab:table4}
are in a logarithmic scale with respect to hydrogen, namely:
log\,$\epsilon$(X)\,=\,log\,[N(X)/N(H)] + 12.0. The abundances of the elements
relative
to hydrogen and iron are expressed as [X/H]\,=\,log\,$\epsilon$(X)$_{star}$\,-\
log\,$\epsilon$(X)$_{sun}$  and [X/Fe]\,=\,[X/H]\,-\,[Fe/H] respectively.

\subsection {Uncertainties in the elemental abundances}
\label{sec:uncert}

In this work the atomic data as the harmonic oscillator strength log\,$gf$ and low
excitation potential $\chi$ (eV) were collected from different sources, e.g.
\cite{RaoEtAl2012} and references therein. The errors in the atomic data
vary
from element to element, for example the accuracy of experimental values for
\ion{Fe}{i} 
and \ion{Fe}{ii} may be between 5\% and 10\%. For other Fe-peak elements, errors in
their
log\,$gf$ may range between 10\% and 25\%, while for neutron-capture elements the
precision 
is about 10\%.

In order to estimate the uncertainties in the resulting elemental abundances, we
calculated the effect of varying temperature, gravity and turbulent velocity within
the ranges
$\pm$250K, $\pm$0.25 dex and $\pm$0.5 kms$^{-1}$ respectively. The sensitivity of the
derived 
abundances to these changes in the model atmosphere input parameters is exemplified
in 
Table~\ref{tab:table5}.
The errors due to equivalent widths are random ($\sigma_{ran}$) since they depend on
several 
factors such as the position of the continuum, the signal-to-noise (S/N) ratio and
spectral type of the star. In contrast, the errors due to the atmospheric parameters
and atomic 
data are systematic ($\sigma_{sys}$) and depend on the precision with which we have
been able 
to estimate them.

The total error $\sigma_{tot}$ for each element can be estimated through the square
root 
of the quadratic sum of the random and systematic errors (see Table~\ref{tab:table5}).

\section{INDIVIDUAL ABUNDANCES}
\label{sec:discuss}

\subsection{IRAS\,02528\,+\,4350}
\label{sec:iras02528}

 The spectrum has broad and very low-intensities 
lines and their cores are affected by noise. We report a rotation velocity of 
about 55 km s$^{-1}$ and a heliocentric radial
velocity of $-$0.9$\pm$0.3 km s$^{-1}$ derived from \ion{Na}{i}\@ D$_{1}$ and D$_{2}$
measured between two subsequent orders.

The star is moderately metal-poor and has few identified elements. CNO abundances are 
based upon the synthesis of two C lines ($\lambda$5052\AA and $\lambda$5830\AA)
and the \ion{O}{i} triplet at $\lambda$6150\AA. We derive a moderate enrichment LTE
value of carbon
[C/H] of $-$0.13 and [C/Fe] of $+$0.77, while the oxygen shows a similar enrichment
to that of carbon, [O/H]\,=\,$-$0.09 and [O/Fe]\,=\,$+$0.81 respectively.
Non-LTE corrections of $-0.39$, $-0.17$ and $\sim0.25$ for C, O and Fe
were taken from \cite{Venn1995}, \cite{TakedaTakada1998} 
and \cite{LindEtAl2012} leading to values of [C/Fe]\,=\,$+$0.13
and
[O/Fe]\,=\,$+$0.39. The non-LTE corrected values of carbon and oxygen abundances
indicate a C/O ratio
slightly lower than the solar value (C/O$\sim$0.5), i.e. $+$0.3.

The Na abundance is derived from the line at $\lambda$6160 and shows a moderate
enhancement,
[Na/Fe]\,=\,$+$0.79. A non-LTE correction of $-$0.10 taken from \cite{LindEtAl2011}
leads to a value of [Na/Fe]\,=\,$+$0.44 which indicates that such enrichment is real
and is possibly caused by proton capture on $^{22}$Ne in H-burning region.
The behaviour of $\alpha$-elements ([$\alpha$/Fe]\,=\,$+$0.14) is not fully compatible
with the thick
disc values, in fact, [Ca/Fe]\,=\,$+$0.32 shows a value typical of thick disc stars
but
[Mg/Fe]\,=\,$+$0.08, [Si/Fe]\,=\,$+$0.19 and [Ti/Fe]\,=\,$-$0.05 are lower than
expected
to thick disc population. On the other hand, the Cr, Sr and Ba abundances [X/H]
show depletion, and unfortunately 
S and Zn elements are not present to corroborate this effect. In addition,
Mg, Si and Cr do not show the trend of dust-gas winnowing (DG-effect).

\subsection{IRAS\,05338\,-\,3051}
\label{sec:iras05338}

IRAS\,05338\,-\,3051 is a semi-regular variable with a pulsation period of 105.7 days and 
$V$ light variations between 9.30 and 10.30 mag \citep []{KukarkinEtAl1971, SamusEtAl2009}. 

The heliocentric radial velocity of the star measured from the spectrum taken on Feb 10, 
2017 was of $+$5.2$\pm$0.5 kms$^{-1}$.
The spectrum of the star is not crowded but contains a large number of lines for several 
important elements. 
The star is metal-deficient with [Fe/H]=$-$1.3 dex. The high galactic latitude of 
IRAS\,05338\,-\,3051 combined with its low
metallicity indicates that the star probably belongs to the old-disk population of 
the Galaxy.

In view of the fact that our object is cooler than the Sun, we use the Arcturus
spectrum as a reference
to calibrate our line list of the identified EWs. The condition for verifying the 
quality of our line list requires that the abundances obtained agree satisfactorily
with those estimated 
in other works. By comparing our abundances with those obtained by \cite{RamirezAllende2011}, 
we noted no systematic differences greater than 0.03 dex. The elemental abundances of 
IRAS\,05338\,-\,3051 are presented in Table~\ref{tab:table3}.

IRAS\,05338\,-\,3051 does not show sodium enrichment ([Na/Fe]\,=\, $-$0.15 dex). A
moderate enrichment
of the $\alpha$ elements (Mg, Si, Ca, Ti) of $+$0.33 dex has been observed, typical of
the thick disc population. Magnesium ([Mg/Fe]\,=\, $+$0.37) and silicon ([Si/Fe]\,=\,
$+$0.60)
show much larger values than calcium ([Ca/Fe]\,=\, $+$0.13) and titanium ([Ti/Fe]\,=\,
$+$0.23).
Iron-peak elements (Sc, V, Cr, Ni) show [X/Fe] abundances mildly enhanced that vary
from
$+$0.05 to $+$0.46 dex. We observe also that the vanadium and nickel are surprisingly
overabundant 
relative to the solar value, [V/Fe]\,=\, $+$0.46 and [Ni/Fe]\,=\, $+$0.35 dex.

Neutron-capture (s-and-r process) elements have [X/Fe] abundances moderately enriched.
s-process elements as Y, Zr, Ba, La, Ce, Nd and Sm show abundances varying from
$+$0.19
to $+$0.77 dex, while r-process elements as Pr, Eu range from $+$0.75 to $+$0.89 dex
respectively.
Our analysis covers light s-process elements (ls) Y and Zr and heavy s-process
elements (hs)
Ba, La, Ce and Nd. We derive a mean [ls/Fe] of $+$0.53 dex and mean [hs/Fe] of $+$0.47
dex. 
That leads to the ratio of hs-elements to ls-elements [hs/ls] of $-$0.06 dex.

With a moderate enrichment of s-elements one would expect that C is enriched,
however, 
surprisingly we note that it is very deficient ([C/H]\,=\,$-$1.13 dex).
Our C abundance has been estimated by synthesizing several regions with C$_{2}$
($\lambda$4732--4744; $\lambda$5133--5165; $\lambda$5626--5636) and CH bands,
($\lambda$4290--4310). The N abundance is derived from the CN region at
$\lambda$4213--4216. We
observe a moderate enhancement of [N/Fe] abundance of $+$0.69 which involves a
conversion
of initial C into N through the CN-cycle. The O abundance was estimated from three
lines 
($\lambda$5577, $\lambda$6300 and $\lambda$6363) and shows an enriched value of 
[O/Fe]\,=\,$+$0.84 dex. The C/O ratio is about of $+$0.12.

Our derived abundances have the same trend observed in 
V453 Oph studied by \cite{DerooEtAl2005}. V453 Oph is a
RV Tauri star which shows low metallicity ([Fe/H]\,$\sim$ $-$2.2
dex), s-process enrichment but C deficiency and has no infrared
excess.
 
\subsection{IRAS\,17279\,-\,1119}
\label{sec:iras17279}

IRAS\,17279\,-\,1119 is a PAGB star that has been the subject of
several previous abundance 
analysis. \cite{VanWinckel1997} suggested that it is a pulsating
variable. \cite{Bogaert1994} identified two possible periods; 61 and 93 day, 
although more photometric data would be required to confirm them.

\cite{VanWinckel1997} derived the values $T_{\rm eff}$\,=\,7400K, log\,$g$=0.5 and [Fe/H]\,=\,$-$0.7 dex. 
From a detailed abundance analysis the star is revealed as metal-poor,
with C, N moderately enriched and enriched s-process elements.

\begin{figure}
\begin{center}
  \includegraphics[width=8.cm,height=8cm]{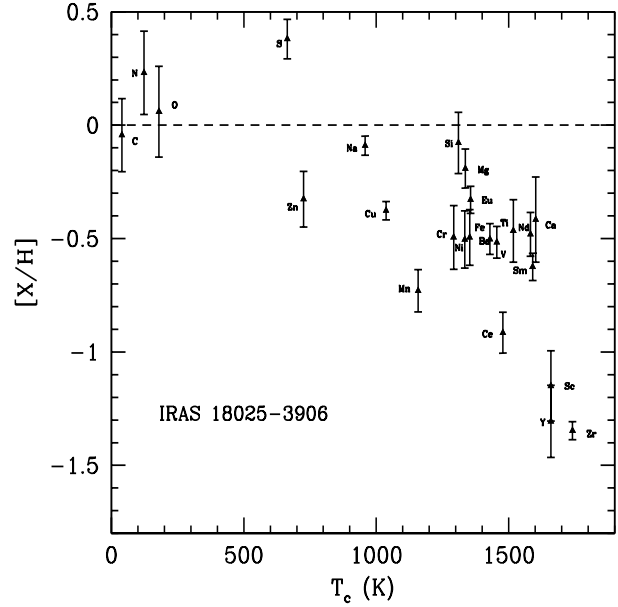}
  \caption{Abundance [X/H] versus condensation temperature T$_{C}$
for IRAS\,18025\,-\,3906.}
  \label{fig:figure5}
\end{center}
\end{figure}

\cite{ArellanoEtAl2001} estimated  $T_{\rm eff}$\,=\,7300K, log\,$g$=1.5, $\xi_{t}$\,=\,4.6 km
s$^{-1}$ 
and [Fe/H]\,=\,$-$0.6 dex.
They also found that the star is metal-poor, and have a moderate enrichment in C, Na,
Sc and 
s-process elements. In agreement with the above results \cite{RaoEtAl2012} estimated $T_{\rm eff}$\,=\,7250K, log\,$g$=2.25, $\xi_{t}$\,=\,4.7 km s$^{-1}$ and  [Fe/H]\,=\,$-$0.43 dex. 

We note that the metallicity estimated by \cite{RaoEtAl2012}
is slightly higher 
than those obtained by \cite{VanWinckel1997} and \cite{ArellanoEtAl2001}.
The present N abundance, [N/Fe]\,=\,$-$0.3 dex, is higher than that obtained by \cite{VanWinckel1997} ([N/Fe]\,=\,$-$0.6 dex). C and s-process abundances are moderately enriched (see Table 8).

A recent abundance analysis was carried out by \cite{DeSmedtEtAl2016}. Their
derived atmospheric parameters were $T_{\rm eff}$\,=\,7250K, log\,$g$=1.25,
$\xi_{t}$\,=\,3.0 km s$^{-1}$ and [Fe/H]\,=\,$-$0.64 dex.
According to these authors, IRAS\,17279\,-\,1119 has a moderate C enhancement,
displays
a mild s-process enrichment and most of the s-process elements have an [X/Fe]\,$<$ 1.0
dex.
Their calculation of [N/Fe]\,=\,$-$0.07 dex, does not shows the enhancement obtained by \cite{RaoEtAl2012}, (+0.63).
This object
shows 
evidence of being moderately metal-poor and have moderately enriched abundances of
s-process elements . 

For IRAS\,17279\,-\,1119 we have adopted the atmospheric parameters 7250K, 1.25, 
4.4 km s$^{-1}$ and $-$0.59 dex for $T_{\rm eff}$, log\,$g$, $\xi_{t}$ and [Fe/H]
respectively 
(see Table~\ref{tab:table2}). 
We have determined the carbon, nitrogen and oxygen abundances for this
object based upon a fairly large number of lines. The abundances [C/H] and
[C/Fe] of $-$0.18 and 
$+$0.41, [N/H] and [N/Fe] of $+$0.17 and  $+$0.76 and [O/H] and [O/Fe] of $-$0.36 and
$+$0.23 were 
found. At this temperature, the non-LTE corrections are $-$0.17, $-$0.30
and $-$0.10 for C, N and O respectively and
imply [C/Fe] of $+$0.10, [N/Fe] of $+$0.30 and [O/Fe] of $-$0.03. The moderate
N enrichment, implying conversion of initial carbon into N through CN-cycle, and the
observed
near-solar O abundance, does not indicate ON processing. The C/O ratio is about 0.82
dex.

IRAS\,17279\,-\,1119 shows relative enrichment of sodium possibly caused by proton
capture on
$^{22}$Ne. The $\alpha$-elements display a consistent enrichment relative to Fe of
$+$0.3 dex
typical for the thick disc stars. The Fe-peak elements [Ni/Fe] and [Zn/Fe] show a
moderate
enrichment of $+$0.38 and $+$0.49 dex respectively.
s-process elements exhibit a mild enhancement observed previously by \cite{VanWinckel1997}, \cite{ArellanoEtAl2001}, \cite{RaoEtAl2012} and \cite{DeSmedtEtAl2016}.
We estimated the heavy s-process [hs/Fe]\,=\,$+$0.94 and the light s-process
[ls/Fe]\,=\,$+$0.91
while the [hs/ls] ratio is $\sim$0.03. The abundances of IRAS\,17279\,-\,1119
are given in Table~\ref{tab:table3}.

Our abundances are in agreement, on average within $\pm$0.2 dex, with those
derived by \cite{DeSmedtEtAl2016} with the exception of nitrogen and manganese. Our
N
abundance ([N/Fe]\,=\,$+$0.76 dex) however has a similar value 
to that obtained by \cite{RaoEtAl2012} of [N/Fe]\,=\,$+$0.63 dex, while Mn
abundance is very deficient, [Mn/Fe]\,=\,$-$0.59 dex. Manganese deficiency is
difficult to explain. According to \cite{DeSmedtEtAl2016}, IRAS\,17279\,-\,1119 is the first s-process
rich PAGBs discovered to be in a spectroscopic binary and the precursor of
extrinsically s-process
enriched stars.

\subsection{IRAS\,18025\,-\,3906}
\label{sec:iras18025}

IRAS\,18025\,-\,3906 (GLMP 713) has been classified as G2 by
\cite{HuEtAl1993}. The heliocentric radial velocity from 86
lines in the spectrum taken on September 30, 1998 is
$-$120.2$\pm$0.6 kms$^{-1}$. Our value of LSR velocity ($-$113.0
kms$^{-1}$) is very similar that
obtained by \cite{SilvaEtAl1993} of $-$116.0 kms$^{-1}$. 
The photospheric abundances of IRAS\,18025\,-\,3906 show a
moderate metal-deficiency ([Fe/H]\,=\,$-$0.50 dex). This star
exhibits a large number of \ion{C}{i} lines ($\sim$ 20 lines).
The N abundance appears enriched by the CN process
([N/Fe]\,=\,$+$0.74), while [C/Fe]\,=\,$+$0.46 points to the
products of He burning being brought to the
surface. The [O/Fe] of $+$0.56 is slightly larger than [C/Fe]
but the C/O ratio is $\sim$0.43.
The Na abundance has a moderate enrichment probably caused by
proton capture on $^{22}$Ne ([Na/Fe]\,=\,$+$0.41).
A [$\alpha$/Fe]\,=\,$+$0.22 ratio and the high radial velocity,
are similar to the  observed in objects of the thick disk. 
Individually, S is enriched ([S/Fe]\,=\,$+$0.88), whereas the Mg
and Si have a moderate enrichment ([Mg/Fe]\,=\,$+$0.31 and
[Si/Fe]\,=\,$+$0.43). The Zn abundance is nearly solar
([Zn/Fe]\,=\,$+$0.18). Fe-peak abundances [X/Fe] varying between
$-$0.65 to $+$0.01, while the s-process elements
range from $-$0.84 to $+$0.18. We have plotted in
Fig.~\ref{fig:figure5} the observed abundances for
IRAS\,18025\,-\,3906 as function of the condensation temperature
for the elements \citep []{Lodders2003}. From the relations
[X/H] versus T$_{c}$ we noted that IRAS\,18025\,-\,3906 shows a
modest depletion. 

To verify if this object actually presents depletion, we
calculated the abundances of photospheric depletion traces, such
as [Zn/Ti] and [S/Ti] respectively. For
IRAS\,18025\,-\,3906 we found [Zn/Ti]\,=\,$+$0.15 and 
[S/Ti]\,=\,$+$ 0.86, showing that
this object is not depleted. In addition, its SED has shell-type
structure \citep{HuEtAl1993} suggesting the presence
of cold dust away from the central star and its non-binary
nature. There is observational
evidence that PAGBs with a C/O $<$ 1 ratio are rich in oxygen
and do not exhibit s-process enrichment, which suggests that
the TDU has not been efficient in the AGB phase. The C/O
$\sim$0.43 and the non-enrichment of s-process elements could
point that IRAS\,18025\,-\,3906 is probably a PAGB with
this type of peculiarities.

However, one of the problems that difficult to predict
the PAGB stage for optically visible galactic PAGBs is that
their distances and luminosities are unknown. Gaia DR2 source
provides parallaxes for a $\sim$1.7 billon of stars in the Milky
Way, although not all the stars present in this source have a
precise parallaxes ($\sigma_{p}$/$p$ $\lesssim$ 20\%). For some
stars with distances greater
than 4 kpc from the Sun, the relative error could exceed 25\%.
We can observe that the parallax for IRAS\,18025\,-\,3906 has
a very large relative error, close to 99\%
($p$\,=\,0.4399$\pm$0.4387 mas). \cite{VickersEtAl2015} assume a
luminosity of 6000$\pm$1500 L$_{\odot}$ for this object, and
indicate that it belongs to the intermediate thin disk with
ranges in age, metallicity and mass of 0.7--3.0
Gyr, $-$0.2 to $+$0.5 and 1.6--3.0 L$_{\odot}$. However, we can
see that our metallicity is lower than the proposed range for this population ($-$0.50 dex), this would lead
us to deduce that IRAS\,18025\,-\,3906 rather belongs to the old
thin disk population with a luminosity of 3500$\pm$1500
L$_{\odot}$ (see Table 3 in \cite{VickersEtAl2015}).

\subsection{IRAS\,18386\,-\,1253}
\label{sec:iras18386}

This PAGB candidate was analysed by \cite{PereiraMiranda2007}
with low-resolution spectra and classified it as a probable F5
type star. The heliocentric radial velocity obtained from 115
lines in the spectrum taken on July 29, 2009 is $+$83.2$\pm$0.5
kms$^{-1}$. The star shows a marginal iron deficiency,
[Fe/H]\,=\,$-$0.18 dex. 

\begin{figure}
\begin{center}
  \includegraphics[width=8.cm,height=7cm]{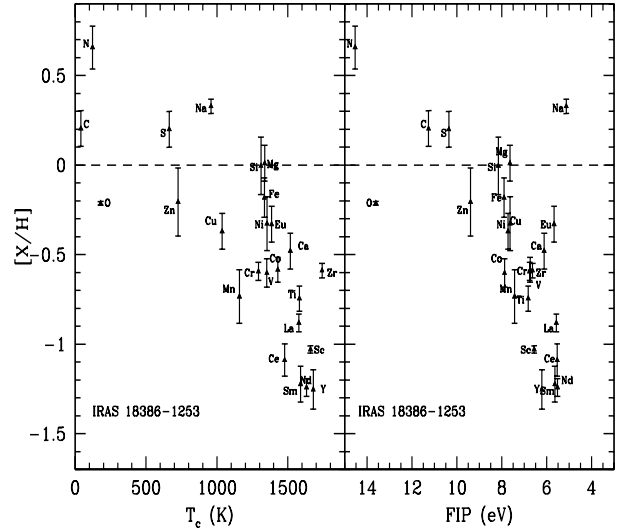}
  \caption{Abundance [X/H] versus condensation temperature
  T$_{C}$ (left panel)
and first ionization potential FIP (right panel)
for IRAS\,18386\,-\,1253.}
  \label{fig:figure6}
\end{center}
\end{figure}

The C abundance [C/Fe]\,=\,$+$0.39 dex, derived from a large
number of lines shows a moderate enrichment and [N/Fe]=$+$0.84
dex indicates strong signatures of the CN cycle.
The [\ion{O}{i}] forbidden lines $\lambda$5577\AA\@ and
$\lambda$6363\AA\@ lead to [O/Fe]\,=\,$-$0.03.
IRAS\,18386\,-\,1253\@ is C-rich since the C/O ratio is slightly
greater than one (C/O $\sim$1.4).
The s-process elements do not display any enrichment.
[$\alpha$/Fe]\,=\,$+$0.25 shows a consistent enrichment relative
to Fe expected for the thick disk population but with a low 
metallicity. The photospheric depletion tracers 
[Zn/Ti]\,=\,$+$0.54 and [S/Ti]\,=\,$+$0.95 indicate
that IRAS\,18386\,-\,1253 is mildly depleted.

\begin{figure}
\begin{center}
  \includegraphics[width=8.cm,height=8cm]{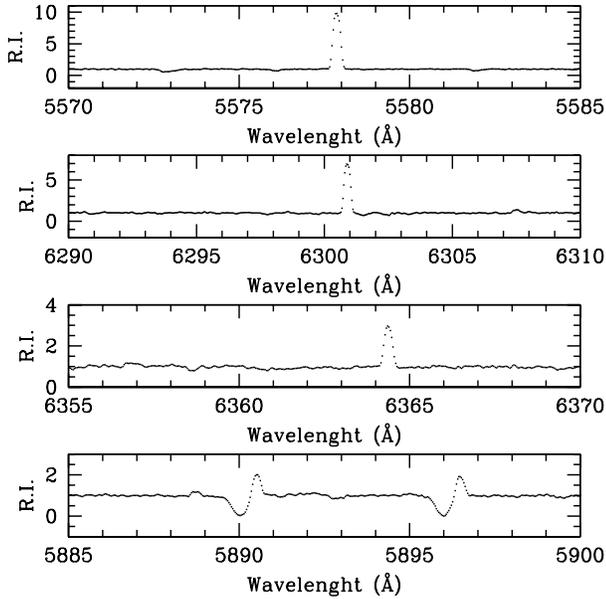}
  \caption{Emissions present in the forbidden oxygen lines at $\lambda$5577\AA,
$\lambda$6300\AA\@ and
$\lambda$6364\AA\@ and \ion{Na}{i}\@ D doublet for IRAS\,20259\,+\,4206.}
  \label{fig:figure7}
\end{center}
\end{figure}

It is tempting to ascribe this depletion only to dust
condensation, however the large scatter observed may suggest
that another mechanism operates simultaneously.
This other mechanism might be the first ionization potential
(FIP) effect initially seen in the solar chromosphere and corona
where ions of low-FIP elements rather than neutral
atoms are fed from the chromosphere to the corona. In this sense
the plot [X/H] versus FIP in Fig.~\ref{fig:figure6}
shows that there is a good correlation between the high-FIP and
low-FIP elements with the exception of O and Na for
IRAS\,18386\,-\,1253. This phenomenon was observed by
\cite{RaoReddy2005} for the two RV Tauri stars CE Vir and EQ
Cas. These authors propose that singly ionized elements escaped
as stellar wind rather than being coupled to the radiation
pressure on the dust.

It is clear that elements with a FIP of 8\,eV or lower
show systematic depletion. This phenomenon has also been
observed in the stars CpD\,-\,62\,5428 \citep
[]{GiridharEtAl2010} and IRAS\,18321\,-\,1401 \citep
[]{MolinaEtAl2014}. The smooth anti-correlation between the
abundances and the FIP effect suggests that the removed elements
systematically correspond to species with the first ionized
states.

\cite{GarciaSegura2005} suggest that magnetic field
driven winds have the potential to remove ionized elements in
the photosphere of PAGB stars although the existence of stellar
winds must be verified for IRAS\,18386\,-\,1253. The mechanism
that provides the overionization, however, is not  entirely
understood. A multi-wavelength monitoring of this star
IRAS\,18386\,-\,1253 may be required to understand its peculiar
abundance patterns.

\subsection{IRAS\,20259\,+\,4206}
\label{sec:iras20259}

The present abundance analysis of IRAS\,20259\,+\,4206 is the
first performed for this star. For this object a heliocentric
radial velocity of $-$15.6$\pm$0.6 kms$^{-1}$ was
measured using 125 lines. The adopted model atmosphere
parameters are given in the Table~\ref{tab:table2} and
the resulting abundances are listed in Table~\ref{tab:table4}.
The profiles of the \ion{Na}{i}\@ D lines show emission in the
right wings, suggesting that this object undergoes a strong mass-loss (see Fig.~\ref{fig:figure7}). 

\begin{figure}
\begin{center}
  \includegraphics[width=8.cm,height=8cm]{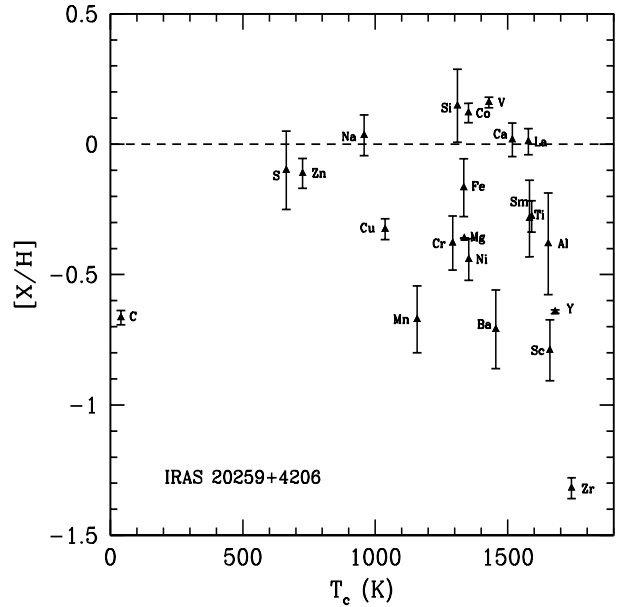}
  \caption{Abundance [X/H] versus condensation temperature
  T$_{C}$
for IRAS\,20259\,+\,4206.}
  \label{fig:figure8}
\end{center}
\end{figure}

The star is slightly metal poor ([Fe/H]\,=\,$-$0.17 dex). It also shows an under abundance of carbon ([C/H]\,=\,$-$0.64) obtained from two lines; $\lambda$6587.6\AA\@ and $\lambda$5380.3\AA.
Due to our limitation in the spectral range we can not measure
the abundance of N, therefore we can not predict the CN-cycle
efficiency. We also can not measure the O abundance. The
\ion{O}{i} triplet lines at $\lambda$6155--60\AA\@ are
contaminated while the forbidden lines [\ion{O}{i}]
$\lambda$5577\AA\@, $\lambda$6300\AA\@ and
$\lambda$6364\AA\@ are affected by emission (see
Fig.~\ref{fig:figure7}). 
Sodium has a modest enhancement ([Na/Fe]\,=\,$+$0.20).
The [$\alpha$/Fe] ratio of $\sim$0.06 obtained from S, Si and
Mg, is similar to those found in thin disk objects.
Sulfur abundance is derived from synthesis by taking into
account only the line $\lambda$6757.1\AA\@ of the triplet at
$\lambda$6743-58\AA. The $\lambda$6743\AA\@ and
$\lambda$6748\AA\@ lines could be affected by blends. Its value
is near to solar, i.e. [S/Fe]\,=\,0.07 dex.

In Fig.~\ref{fig:figure8} we have plotted the observe abundances for IRAS\,20259\,+\,4206 as function of  the condensation temperature.
Modest depletion patterns with great scatter are observed in
certain elements with T$_{c}$ $>$ 1300K, e.g. Cr, Mg, Fe, Ti,
Sm, Al, Ni, Ba, Sc and Y, while Zr is more depleted by at least
one order of magnitude. Other elements such as Si, Ca, V, Co, La
do not show depletion. The low signal-to-noise (S/N $\sim$30),
the moderate resolution and the limitation on the spectral 
range make necessary that this object is analyzed with more favourable data.

\section{Discussion}
\label{sec:discuss}

In the sample of PAGB candidate stars in the present study, there are five objects not previously studied spectroscopically; IRAS\,02528\,+\,4350, IRAS\,05338\,-\,3051, IRAS\,18326\,-\,1253, IRAS\,18025\,-\,3906, and
IRAS\,20529\,+\,4206. On the other hand, the elemental abundances of the star IRAS\,17279\,-\,1119, have
been calculated by several investigators \citep []{DeSmedtEtAl2016, RaoEtAl2012, ArellanoEtAl2001, VanWinckel1997}. An exhaustive abundance analysis is carried out to discriminate among the enormous chemical variety that these kind of objects may exhibit.

In the majority of the sample stars we find evidence of CN processing, with
the exception of IRAS\,02528\,+\,4350 and IRAS\,20259\,+\,4206, 
in which it was not possible to determine the N abundance. 
In the remaining objects, the [N/Fe] ratio ranges between $+$0.7 and $+$0.8 dex. 
We have also seen the signature of mixing of He-burning products via the TDUP for IRAS\,02528\,+\,4350, IRAS\,05338\,-\,3051,
IRAS\,17279\,-\,1119, IRAS\,18025\,-\,3906 and
IRAS\,18326\,-\,1253. In fact, the [C/Fe] ratio ranges from $+$0.10 to $+$0.46 dex. 
IRAS\,20259\,+\,4206, clearly exhibits an underabundance of carbon of $-$0.50 dex. This deficiency of C is greater than that caused by the CN-cycle, which leads to a N enhancement. A possible alternative to the observed carbon reduction could point out towards a Hot Bottom Burning (HBB) phenomenon, that operates in AGB stars more massive than 4 M$_{\odot}$, and that
partially converts the C excess in the envelope into N. Unfortunately we have been unable to determine the expected Li enrichment in the case of HBB \citep []{SackmannBoothroyd1992}.

\subsection{Sodium, $\alpha$-elements and s-process elements}
\label{sec:sodium}

For all the sample stars the [Na/Fe] abundance varies between $-$0.15 and
$+$0.79 dex. [Na/Fe] abundance is commonly determined from
lines at $\lambda$5682\AA, $\lambda$5688\AA, $\lambda$6154\AA\@ and
$\lambda$6160\AA.
For the temperature range of our sample (i.e. 4250K--7900K), the non-LTE correction
could vary between 
$-$0.10 to $-$0.16 dex \citep []{LindEtAl2011}. With this correction, the observed values
are not significantly
affected, which suggests that Na enrichment is real and probably comes via proton 
capture on $^{22}$Ne. Only IRAS\,05338\,-\,3051 shows a subsolar value;
[Na/Fe]\,=\,$-$0.25 dex.

Regarding the $\alpha$-elements, in 
IRAS\,18386\,-\,1253, the [$\alpha$/Fe] was obtained exclusively from Mg,
Si and S because
Ca and Ti could be affected by depletion and non-LTE effect.
IRAS\,18386\,-\,1253 has a value of $+$0.26 typical of thick-disk with
metallicities nearly solar. The high radial velocity observed for this object would confirm
this fact. IRAS\,18025\,-\,3906 exhibits a moderate enrichment of $\alpha$-elements
([$\alpha$/Fe]\,$\sim$\,$+$0.22) expected for objects of the thick-disk.

On the other hand, IRAS\,05338\,-\,3051, exhibits a moderate enrichment of $\alpha$-elements of $+$0.33 derived from Mg, Si, Ca and Ti respectively. Its low-metallicity, [Fe/H]\,=\,$-$1.3, points it as thick-disk or Galactic halo object. The [$\alpha$/Fe] for IRAS\,02528\,+\,4350 of $+$0.14 and its [Fe/H]\,=\,$-$0.9 support an unevolved object of thin-disk population. IRAS\,17279\,-\,1119 shows a moderate enrichment of $\alpha$-elements of $+$0.33 with [Fe/H]\,=\,$-$0.6 suggesting that it is an object of the thick-disk population. IRAS\,20259\,+\,4206 exhibit [$\alpha$/Fe]\,=\,$+$0.05 with [Fe/H]\,=\,$-$0.2 which is typical of objects of the thin-disk.

\begin{figure}
\begin{center}
  \includegraphics[width=8.cm,height=8cm]{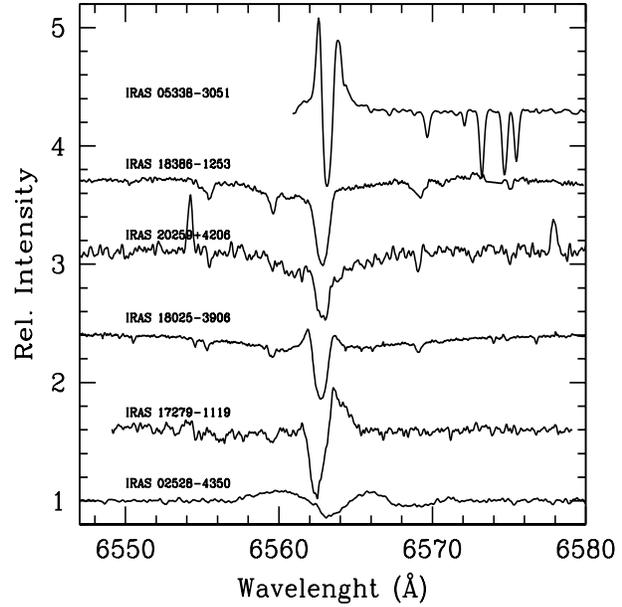}
  \caption{Variability of the absorption and emission H$\alpha$ profiles for
IRAS\,02528\,+\,4350,
IRAS\,05338\,-\,3051, IRAS\,17279\,-\,1119,
IRAS\,18025\,-\,3906, IRAS\,18386\,-\,1253 and IRAS\,20259\,+\,4206.}
  \label{fig:figure9}
\end{center}
\end{figure}

\begin{figure}
\begin{center}
  \includegraphics[width=8.cm,height=8cm]{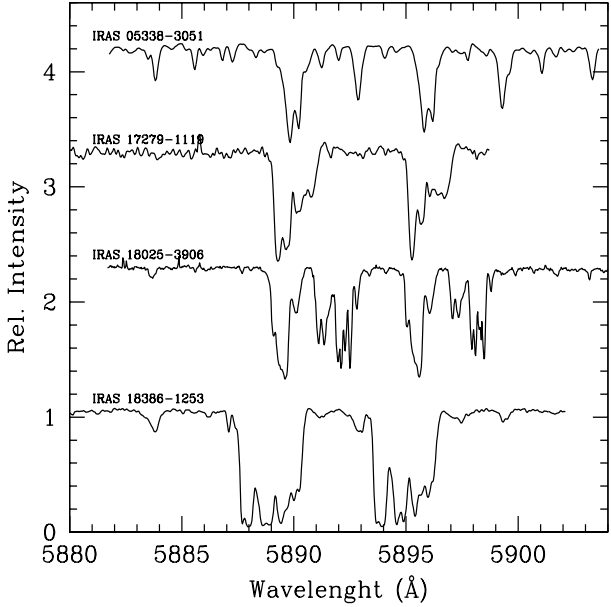}
  \caption{Complex structure of D$_{2}$ ($\lambda$ 5890) and D$_{1}$
($\lambda$ 5895) \ion{Na}{i} line profiles in four objects in the sample. 
Multi-component interstellar absorption lines are present in their spectra.}
  \label{fig:figure10}
\end{center}
\end{figure}

\subsection{H$\alpha$ and Sodium doublet profiles}
\label{sec:halfa-NaID}

The H${\alpha}$ profiles in PAGB stars are affected by instability of the dynamic
processes in the
extended atmosphere and its envelope. These profiles show a range in the
emission-absorption structure
related to the complex atmospheric motions (or perturbations) that propagate
throughout the atmospheric 
layers.

\begin{figure}
\begin{center}
  \includegraphics[width=8.cm,height=8cm]{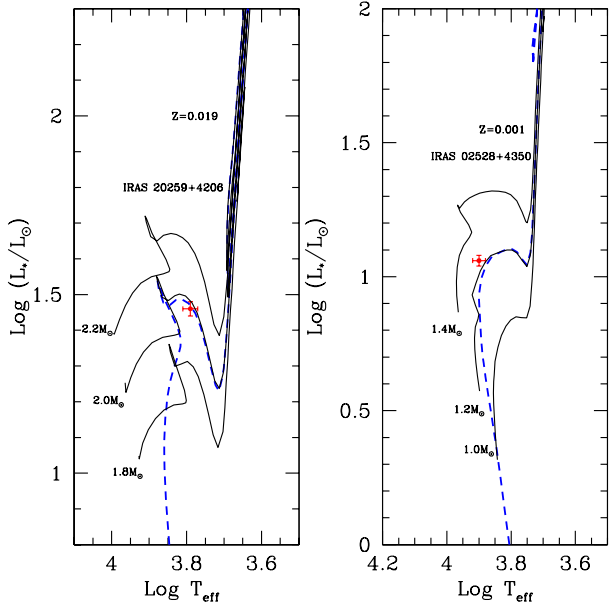}
  \caption{Evolutionary tracks in the HRD for two compositions
[Z\,=\,0.019, Y\,=\,0.273] and [Z\,=\,0.001, Y\,=\,0.230],
where Y and Z represent the helium and metal mass fractions.
IRAS\,02528\,+\,4350 is moderately iron-deficient and
IRAS\,20259\,+\,4206 shows nearly solar metallicity.
The left panel shows the position of IRAS\,20259\,+\,4206 as
compared with the models for Z\,=\,0.019 and masses of 1.8, 2.0
and 2.2 M$_{\odot}$ from \cite {GirardiEtAl2000}. 
The blue dashed line is an isochrone of 1.12 Gyrs. 
The right panel is shows the position of IRAS\,02528\,+\,4350
along with the models for Z\,=\,0.001 and masses 0.8, 1.0 and
1.2 M$_{\odot}$, the shown isochrone is of 3.16 Gyrs.}
\label{fig:figure11}
\end{center}
\end{figure}

Fig.~\ref{fig:figure9} shows the H${\alpha}$ profiles of the six
program stars. In IRAS\,20259\,+\,4206 and IRAS\,18386\,-\,1253,
the H${\alpha}$ profile shows a broad shallow component and
central filling with asymmetric emission. Evidently,
emission is not strong enough to rise above the continuum. This
emission could be produced in a circumstellar cloud of (hot) gas
or an extended envelope \citep []{ArellanoFerro1985}.

IRAS\,05338\,-\,3051, IRAS\,18025\,-\,3906 and
IRAS\,02528\,+\,4350 have a central
absorption flanked by emission peaks on either side displaying
each object a weak, strong and moderate P-Cygni structure. 
Their H${\alpha}$ profiles are very different from star to star.  
In IRAS\,02528\,+\,4350 the line is almost completely filled and
is asymmetric with two emission components in the wings
indicative of a complex, non-spherical
circumstellar shell, e.g. a circumstellar disc \citep []{Klochkova2014}. Similar profiles are seen in the
stars V5112Sgr and
BD\,48$^{0}$\@1220$=$IRAS\,05040\,+\,4820, as shown in figures \,2\@ and 4\@ of \citep []{Klochkova2014}. 
The complex H${\alpha}$ profile of IRAS\,17279\,-\,1119 shows a strong P-Cygni
structure. The presence 
of emission features have been interpreted as high mass loss rate \citep []{TramsEtAl1989}.

Fig.~\ref{fig:figure10} shows the complex structure observed in the \ion{Na}{i}
D$_{2}$
($\lambda$5890) and D$_{1}$ ($\lambda$5895) absorption profiles. We noted that
IRAS\,17279\,-\,1119, 
IRAS\,18025\,-\,3906 and IRAS\,18386\,-\,1253 show multi-component interstellar
absorption
lines in their spectra. The doublet \ion{Na}{i}\@ D lines in IRAS\,05338\,-\,3051, on
the other
hand, are thinner and show double components in their core. 

\subsection{Evolutionary stages}
\label{sec:evolve}

In order to good estimates of the
distances (or luminosities), parallaxes with relative error
less than 20\% are required. The Gaia DR2 source provides a
parallax whose relative error does not exceed 7\% only for three
objects of our program stars; IRAS\,02528\,+\,4350,
IRAS\,05338\,-\,3051 and IRAS\,20259\,+\,4206. For
IRAS\,17279\,-\,1119 and IRAS\,18025\,-\,3906 the errors are too
large; 27\% and 99\%, respectively. For
IRAS\,18386\,-\,1253 its parallax is not present.

In terms of evolution, we could distinguish three
groups among our sample stars. The first one, includes the stars
IRAS\,02528\,+\,4350 and IRAS\,20259\,+\,4206. While one of them is moderately iron-deficient and the other show nearly solar
abundances, their heliocentric radial velocities are small, 
$\alpha$-elements have solar values and do not display s-process
enriched indicating that these are Population I objects. 
The observed C abundance under NLTE of $+$0.13 dex seems to indicate that IRAS\,02528\,+\,4350 is evolved and the O
enrichment ($+$0.4 dex) has also been observed for high galactic
latitude objects. Likewise, IRAS\,20259\,+\,4206 shows a strong
deficiency in carbon which indicates that an additional
mechanism to the FDU operates on it. This effect could also
indicate that IRAS\,20259\,+\,4206 is probably evolved. However,
the Gaia parallaxes indicate luminosities of 
1.06$\pm$0.02 (L$_{*}$\,=\,11$\pm$1 L$_{\odot}$) for IRAS\,02528\,+\,4350  and 1.46$\pm$0.02 
(L$_{*}$\,=\,29$\pm$1 L$_{\odot}$) for IRAS\,20259\,+\,4206, which are values of rather unevolved
objects. The effective temperatures are those adopted in
Table~\ref{tab:table2}. Evolutionary model tracks in the H-R
diagram (HRD) for the compositions [Z\,=\,0.019, Y\,=\,0.273]
and [Z\,=\,0.001, Y\,=\,0.230] are shown in Fig.~\ref{fig:figure11}.  The evolutionary tracks and isochrones
are from \cite {GirardiEtAl2000}.
In the left panel we observe that an evolutionary track with
intermediate-mass of 2.0$\pm$0.1 M$_{\odot}$ and an age of 
1.12 Gyr (blue dashed lines) is representative for
IRAS\,20259\,+\,4206. In the right panel a track with
low-mass of 1.2$\pm$0.1 M$_{\odot}$ and an age of 3.16 Gyr 
(blue dashed lines) are specific for IRAS\,02528\,+\,4350.
The HRD of Fig.~\ref{fig:figure11} suggests that both objects
have low-and-intermediate masses that have not reached the RGB. 
The IRAS two colour diagram, on the other hand, shows that both
objects have mid-IR excess at [12]--[25], which could indicate
the presence of cold dust due to an extended circumstellar
envelope and therefore, they could be confused with evolved
objects, i.e. true PAGBs.

In the second evolutionary group we include the stars
IRAS\,17279\,-\,1119 and IRAS\,05338\,-\,3051.
\cite{DeSmedtEtAl2016} have confirmed the binary and PAGB
nature of IRAS\,17279\,-\,1119. In fact, the luminosity derived
from the Gaia parallax leads to a value of 4.18$\pm$0.26. The
PAGB model sequences of \cite{MillerBertolami2016} at Z\,=\,0.01
suggest that this star has initial ZAMS mass of $\sim$ 3
M$_{\odot}$ and a core mass of $\sim$ 0.706 M$_{\odot}$.
IRAS\,05338\,-\,3051, on the contrary, shows peculiarities in
the observed abundances, e. g. overabundance of
s-process elements without an accompaniment of C enrichment.
Moreover, IRAS\,05338\,-\,3051 does not show mid-IR excess
([12]--[25]\,=\,-0.37), the N ($+$0.7) is not high  enough to
explain the C deficiency from the CN cycle and the light and
heavy s-process elements have values of $+$0.47 and $+$0.53
respectively. Similar peculiarities have also been
observed in two previously studied objects;
IRAS\,06165\,+\,3158 \cite []{RaoGiridhar2014} and V453 Oph
\cite []{DerooEtAl2005}. Our abundances are
clearly comparable with the observed abundances for V453 Oph.
This object is a metal deficient pop. II RV Tauri star with
significant enrichment in heavy elements, C deficient and 
without IR excess. The simulations based on AGB s-process model
indicate that C enrichment is inevitable. In this sense, it is
not easy to explain these abundances with the current
nucleosynthesis models. \cite{DerooEtAl2005} propose several
scenarios to explain the peculiar abundances in V453 Oph such
as parental cloud, the enrichment by a binary companion and
intrinsic s-process enrichment by dredge-up but none of these
scenarios explain satisfactorily the observed abundances.
More high quality observations may be required to select a
possible scenario.
Photometrically, \cite{ArkhipovaEtAl2011}
have found that the light curve of IRAS\,05338\,-\,3051
exhibits stable sinusoidal variations with amplitude modulations
and the pulsation period has remained almost stable for
decades. \cite{VickersEtAl2015} have calculated for this object a luminosity of L$_{*}$\,=\,3500$\pm$1500 L$_{\odot}$
which is characteristic of the old thin-disk, with 
age, metallicity and mass in the ranges 3--8 Gyr,
$-$0.7--$+$0.5 and 1.1--1.6 M$_{\odot}$. 
However, we can see that our object is less metallic ($-$1.3
dex) than the metallicity range proposed for the old thin disk
population. This would lead to us to deduce that
IRAS\,05338\,-\,3051 rather has a luminosity of the thick disk Population of L$_{*}$\,=\,1700$\pm$750 L$_{\odot}$
(or 3.23$\pm$0.16). 
The luminosity of IRAS\,05338\,-\,3051 obtained from the Gaia
parallax of
3.12$\pm$0.06 is within the uncertainty of luminosity
of the thick disk Population. The PAGB model sequences of
\cite{MillerBertolami2016} at Z\,=\,0.001 suggest that 
IRAS\,05338\,-\,3051 has initial ZAMS mass of $\sim$0.9
M$_{\odot}$ and a core mass of $\sim$0.533 M$_{\odot}$ and
an age of $\sim$5.01 Gyr. The similarity on observed abundances
between V453 Oph and our object seems to indicate that
IRAS\,05338\,-\,3051 is a probable PAGB. Extensive radial 
velocity monitoring is required to verify the binary nature of
this object.

The third group contains the stars IRAS\,18025\,-\,3906
and IRAS\,18386\,-\,1253. The two objects show mid-IR excess,
their luminosities have not been clearly established  and
they show peculiarities in their abundance patterns, e.g.
IRAS\,18025\,-\,3906 is O-rich and does not show enrichment of
s-process elements and IRAS\,18386\,-\,1253 shows a mildly
depletion of the refractory elements. The presence
of dust around these objects makes difficult the estimation of their distances and/or
luminosities are difficult to estimate, and questions the
PAGB nature. Therefore, an independent determination
of the luminosity is necessary to achieve a better idea of the
true nature of these
objects. The Gaia source does not provide reliable parallax for
neither of the two objects. An alternative to the estimation  of
the distance and/or luminosity might come from 2MASS photometric
calibrations of \cite{Molina2012}. These calibrations lead to the distance of 2805$\pm$363 pc and a
luminosity of 2.30$\pm$0.11 (L$_{*}$\,=\,200$\pm$51
L$_{\odot}$) for IRAS\,18025\,-\,3906 and a distance of
3267$\pm$422 pc and a luminosity of 2.94$\pm$0.11 
(L$_{*}$\,=\,871$\pm$223 L$_{\odot}$) for
IRAS\,18386\,-\,1253 respectively. These luminosities are 
lower than the expected luminosity of the RGB-tip ($\lesssim$2500
L$_{\odot}$) for stars studied in the Magellanic Clouds
(SCM and LMC) \citep{KamathEtAl2014, KamathEtAl2015}.
This suggests that both objects could have an evolutionary fate totally different from the PAGBs.

In the light of the most recent spectroscopic analyzes
in the search of PAGBs candidates in the Magellanic Clouds (SMC
and LMC), a new population of evolved, dusty stars identifies as post-RGB objects has been revealed, with luminosities below
2500 L$_{\odot}$, gravities between 0 to 2, metallicities from
$-$2.5 to $+$0.5 and masses between 0.28 to 0.45 M$_{\odot}$
\cite []{KamathEtAl2014, KamathEtAl2015}.
The post-RGB are peculiar stars that have interrupted their
normal evolution in the RGB phase from a binary mechanism which
causes that their intrinsic properties (pulsation, loss-mass, 
photospheric chemistry, dust-formation, circumstellar
envelope morphology, etc) are altered and evolve away from
the RGB phase. The final result on these binary evolved objects
is the presence of a Keplerian circumbinary disk of gas and
dust. See the work of \cite{KamathEtAl2016} for
more detailed scenarios of binary formation of the post-RGB
objects. Until now, this class of objects has not been clearly
identified in the Galaxy, likely due to the lack of accurate
measurements of their distances and/or luminosities. However, it
cannot be ruled out that some of the galactic PAGB candidates
that have not yet been studied turn out to be post-RGB objects.

We suggest that IRAS\,18025\,-\,3906 and
IRAS\,18386\,-\,1253 are likely members of the post-RGB class.
Their abundance patterns seem to indicate that both objects are
evolved and within dusty circumstellar environments. Clearly,
the luminosities estimated by \cite{Molina2012} are more lower
than those suggested by \cite{VickersEtAl2015} when
considering the two objects as PAGBs (3500$\pm$1500 and
6000$\pm$1500 L$_{\odot}$). This imples that
IRAS\,18025\,-\,3906 and IRAS\,18386\,-\,1253 may have similar
masses and gravities than the known extragalactic post-RGB
objects.

\section{Summary and Conclusions}
\label{sec:suma-con}

Our present investigation of a set of six PAGB candidates, five
not previously explored, led to confirmation of the PAGB
nature of IRAS\,05338\,-\,3051, as well as two likely Galactic
post-RGB candidates such as IRAS\,18025\,-\,3906 and
IRAS\,18386\,-\,1253, from a detailed atmospheric abundance
analysis with high-resolution spectra and their luminosities.
IRAS\,05338\,-\,3051 do not has infrared excess and probably
has a binary nature, and IRAS\,18025\,-\,3906 and
IRAS\,18386\,-\,1253 might be the first Galactic post-RGB objects confused with true PAGBs.

Other objects studied in the sample, IRAS\,02528\,+\,4350
(metal-poor) and IRAS\,20259\,+\,4206 (metal-solar) show signs
of being unevolved objects. 
Although both objects present unconfirmed signs of selective
removal, their low radial velocities, $\alpha$-elements with
abundances nearly solar and lack of s-process elements
enrichment, confirm their unevolved nature.
Their low mass and luminosity and their gravity larger than 2.0, discard them as probable Galactic post-RGB 
and confirm that their evolution does not exceed the
RGB phase.

\section*{Acknowledgments}

We would like to thank Prof. Sunetra Giridhar for useful discussions on abundances and valuable comments on the text, and also for provideding the spectra of IRAS\,02528\,+\,4350 and
IRAS\,17279\,-\,1119 used in this study.
We are grateful to Dr. Anibal Garc\'ia Hern\'andez and Dr. Olga Zamora for taking the spectrum of  IRAS\,20259\,+\,4206 at the Roque de los Muchachos Observatory and to Dr. Bala
Sudhakara Reddy for providing us with the McDonald spectrum of IRAS\,05338\,-\,3051. AAF is indebted to
DGAPA-UNAM (M\'exico) for a grant through project IN104917.
Numerous comments and suggestions made by the anonymous referee are thankfully acknowledged.

\vspace{17.0cm}
\bibliographystyle{Wiley-ASNA}
\bibliography{2810_biblio}
\end{document}